\documentclass[a4paper,twocolumn,11pt,accepted=2026-08-03]{quantumarticle}
\pdfoutput=1
\usepackage[utf8]{inputenc}
\usepackage[english]{babel}
\usepackage[T1]{fontenc}
\usepackage{amsmath}
\usepackage{hyperref}

\usepackage{tikz}
\usepackage{lipsum}

\usepackage{cite}
\usepackage{amsmath,amssymb,amsfonts}
\usepackage{algorithmic}
\usepackage{graphicx}
\usepackage{textcomp}
\usepackage{xcolor}
\usepackage{listings}
\usepackage{subcaption}
\usepackage{caption}
\usepackage{changepage}
\usepackage{booktabs}
\definecolor{comment-text-color}{rgb}{0,0.8,0.6}
\newcommand{\beq}{\begin{equation}}
\newcommand{\eeq}{\end{equation}}
\def\BibTeX{{\rm B\kern-.05em{\sc i\kern-.025em b}\kern-.08em
    T\kern-.1667em\lower.7ex\hbox{E}\kern-.125emX}}

\begin{document}

\title{Quantum Resource Comparison for Two Leading Surface Code Lattice Surgery Approaches}

\author{Tyler LeBlond}
\affiliation{Computational Sciences and Engineering Division, Oak Ridge National Laboratory, Oak Ridge, USA}
\thanks{tyler.leblond@quantinuum.com.}

\author{Ryan Bennink}
\affiliation{Computational Sciences and Engineering Division, Oak Ridge National Laboratory, Oak Ridge, USA}

\maketitle

\begin{abstract}
    Hamiltonian simulation is one of the most promising candidates for the demonstration of quantum advantage within the next ten years, and several studies have proposed end-to-end resource estimates for executing such algorithms on fault-tolerant quantum processors. Usually, these resource estimates are based upon the assumption that quantum error correction is implemented using the surface code, and that the best surface code compilation scheme involves serializing input circuits by eliminating all Clifford gates. This transformation is thought to make best use of the native multi-body measurement (lattice surgery) instruction set available to surface codes. Some work, however, has suggested that direct compilation from Clifford+T to lattice surgery operations may be beneficial for circuits that have high degrees of logical parallelism. In this study, we analyze the resource costs for implementing Hamiltonian simulation using example approaches from each of these leading surface code compilation families. The Hamiltonians whose dynamics we consider are those of the transverse-field Ising model in several geometries, the Kitaev honeycomb model, and the $\mathrm{\alpha-RuCl_3}$ complex under a time-varying magnetic field. We show, among other things, that the optimal scheme depends on whether Hamiltonian simulation is implemented using the quantum signal processing or Trotter-Suzuki algorithms, with Trotterization benefiting by orders of magnitude from direct Clifford+T compilation for these applications. Our results suggest that surface code quantum computers should not have a one-size-fits-all compilation scheme, but that smart compilers should predict the optimal scheme based upon high-level quantities from logical circuits such as average circuit density, numbers of logical qubits, and T fraction.
\end{abstract}

Quantum error correction (QEC) strives to improve the ability of quantum computers to retain their quantum information over time by making quantum memories robust to environmental noise~\cite{terhal2015quantum}. The surface code is the leading candidate quantum error-correcting code (QECC) for planar nearest-neighbor quantum computer architectures such as those based upon superconducting circuits or Majorana zero modes~\cite{acharya2024quantum, google2023suppressing, zhao2022realization, grans2024improved}, and has also sustained interest for architectures with all-to-all connectivity such as those based upon trapped ions and neutral atoms~\cite{yin2025flexion, ruan2025trapsimd, leblond2023tiscc, bluvstein2024logical, viszlai2023architecture}. 
The mapping of quantum circuits onto fault-tolerant quantum computer architectures requires the translation of abstract logic gates into instructions representing fault-tolerant processes available to a QECC.
For surface codes, this translation results in a substantial space-time overhead not only due to the encoding of logical qubits into large grids of physical qubits but also due to the routing of multi-qubit gates, the production and storage of resource states, and the operational overhead of lattice surgery~\cite{kan2025sparo, silva2024optimizing, leblond2024realistic, beverland2022assessing}.
Though substantial progress has been made in the development of surface code compilation schemes to optimize these costs~\cite{molavi2025dependency, hirano2025locality, hamada2024efficient, silva2024multi, beverland2022surface, beverland2022assessing, litinski2022active, chamberland2022universal, litinski2019game}, 
studies that quantitatively compare costs between different compilation approaches for practical quantum circuits are limited.

In this article, we present a comparison between quantum circuit execution resource costs arising from two leading approaches to lattice surgery compilation for surface code quantum computers: direct compilation from Clifford+T circuits to lattice surgery operations (direct Clifford+T) and sequential Pauli-based computation (SPBC), where Clifford operations are eliminated from the circuit at the expense of circuit parallelism. We perform resource analyses for two different implementations of Hamiltonian simulation, namely Trotterization and quantum signal processing (QSP), for a variety of application instances obtained using Los Alamos National Laboratory's Quantum Applications Specifications and Benchmarks tool~\cite{coffrin2023quantum, bartschi2024potential}, and use computation time and space-time footprint as metrics for resource cost comparison. To generate resource estimates in the direct Clifford+T scheme, we use the recently-upgraded open-source Lattice Surgery Compiler~\cite{leblond2024realistic, watkins2023high}, while to generate resource estimates in SPBC we opt for a simpler footprint analysis akin to the original proposal in Ref.~\cite{litinski2019game}. 
While explicit compilation is essential in the direct Clifford+T scheme to accurately evaluate routing and magic state overhead for circuits with varying circuit densities and magic state consumption profiles~\cite{leblond2024realistic}, SPBC can be analyzed without explicit compilation because its resource costs depend only on the number of logical qubits and the number of T gates in the logical circuit~\cite{blunt2024compilation,litinski2019game}. 

For the set of applications we consider, our results suggest that the optimal combination of quantum algorithm (e.g. Trotterization or QSP) and lattice surgery compilation scheme (e.g. direct Clifford+T or SPBC) depends both on the application and the metric (computation time or space-time footprint) to be optimized. However, we have been able to gain insight into trends underlying the fault-tolerance-layer resource estimates (FTRE) of these applications. For instance, we find that both the computation
time and space-time footprint of Trotter-Suzuki circuits
benefit substantially from the direct compilation of Clifford+T gates into parallelizable lattice surgery instructions rather than SPBC, and that this benefit increases as a power law in the number of (circuit-level) logical qubits\footnote{This is complementary to recent results from Ref.~\cite{hirano2025locality}}. The benefit to computation time makes sense with the observation that Trotter-Suzuki circuits have high gate densities and hence more parallelism that can be exploited in direct Clifford+T compilation (as an aside, we find that randomly-generated circuits have a computation time overhead that scales similarly to Trotter-Suzuki circuits.) It was less clear \textit{a priori} that the space-time footprint (a proxy for overall resource costs) for Trotter-Suzuki circuits should also be improved in direct Clifford+T compilation over SPBC, because the higher parallelism for direct Clifford+T compilation requires greater space for magic state distillation and data routing. For the largest application we consider (Trotterized quantum dynamics simulation of the $\mathrm{\alpha-RuCl_3}$ complex), direct Clifford+T compilation utilizes almost 450 magic state factories while SPBC utilitzes only three factories and has a smaller spatial footprint. Nevertheless, the total space-time footprint of direct Clifford+T compilation is 20$\times$ lower than that of SPBC. While the magic state distillation cost is larger under Clifford+T compilation, we find that it does not asymptotically dominate the spatial footprint.

For QSP circuits, on the other hand, we find that the space-time footprint slightly benefits from SPBC. This is intuitive because the logical circuits are mostly serial (average gate density is small and decreases with problem size), thus there is more benefit from Clifford elimination and the smaller layouts available to SPBC. The compilation scheme that minimizes computation time, however, loosely correlates with the T fraction of the circuit.

Interestingly, Trotterization with direct Clifford+T compilation provided optimal resource estimates by both metrics for the largest of the applications we consider ($\mathrm{\alpha-RuCl_3}$), while the QSP algorithm with SPBC led to the lowest total space-time footprint for the smaller applications.  Our results suggest that Trotterization with direct Clifford+T compilation will provide superior space-time footprints for larger applications than the ones considered in this study.

The rest of this paper is organized as follows. In Section~\ref{sec:app} we provide details on the application benchmarks we consider. In Section~\ref{sec:twoapproaches} we explain the assumptions underlying the lattice surgery compilation and resource estimation schemes that we compare. In Section~\ref{sec:subcircuit} we explain how we combine resource estimates for subcircuits into those of whole algorithms. In Section~\ref{sec:results} we discuss our main results. Finally, in Section~\ref{sec:conclusion} we provide concluding remarks.

\begin{table*}[t]
\centering
\caption{Logical-layer resource estimates (LRE) for the application benchmarks under consideration (see Sec.~\ref{sec:app}). For Trotterization, an `occurrence' corresponds to a single Trotter step, while for QSP a step consists of a bundle of key subcircuits.}
\scalebox{0.75}{
    \begin{tabular}{|l|l|rrrrr|rr|}
    \toprule
    Application & Algorithm & \# Occurrences & \# LQ & \# Gates & \# T Gates & Depth & Avg. Circuit Density & T Fraction \\
    \midrule
    TFIM (square) & Trotter & 5.37e+07 & 100 & 7.84e+12 & 3.08e+12 & 1.70e+11 & 4.63e-01 & 3.93e-01 \\
    TFIM (triangle) & Trotter & 2.17e+08 & 1024 & 4.44e+14 & 1.76e+14 & 1.11e+12 & 3.91e-01 & 3.96e-01 \\
    Kitaev (triangle) & Trotter & 4.33e+09 & 1024 & 7.12e+15 & 2.76e+15 & 2.17e+13 & 3.22e-01 & 3.87e-01 \\
    TFIM (cube) & Trotter & 2.76e+08 & 1728 & 9.47e+14 & 3.77e+14 & 1.38e+12 & 4.01e-01 & 3.98e-01 \\
    Kitaev (honeycomb) & Trotter & 2.23e+09 & 2176 & 3.96e+15 & 1.51e+15 & 6.23e+12 & 2.95e-01 & 3.81e-01 \\
    $\mathrm{\alpha-RuCl_3}$ & Trotter & 2.16e+08 & 2176 & 1.79e+15 & 7.01e+14 & 2.00e+12 & 4.14e-01 & 3.91e-01 \\
    \hline
    TFIM (square) & QSP & 2.76e+05 & 119 & 7.57e+10 & 1.57e+10 & 5.60e+10 & 1.19e-02 & 2.07e-01 \\
    TFIM (triangle) & QSP & 4.16e+06 & 1049 & 1.71e+13 & 3.88e+12 & 1.32e+13 & 1.28e-03 & 2.27e-01 \\
    Kitaev (triangle) & QSP & 4.02e+06 & 1049 & 1.47e+13 & 3.60e+12 & 1.16e+13 & 1.24e-03 & 2.46e-01 \\
    TFIM (cube) & QSP & 6.71e+06 & 1755 & 3.09e+13 & 4.67e+12 & 2.08e+13 & 8.93e-04 & 1.51e-01 \\
    Kitaev (honeycomb) & QSP & 4.36e+06 & 2201 & 1.65e+13 & 3.96e+12 & 1.29e+13 & 6.01e-04 & 2.40e-01 \\
    $\mathrm{\alpha-RuCl_3}$ & QSP & 5.53e+07 & 2205 & 9.29e+14 & 2.07e+14 & 7.07e+14 & 6.18e-04 & 2.23e-01 \\
    \bottomrule
    \end{tabular}
}
\label{tab:lre}
\end{table*}

\section{Application Benchmarks}
\label{sec:app}

Los Alamos National Laboratory (LANL) has curated a set of applications for quantum computers that are expected to have high scientific utility~\cite{bartschi2024potential} and has published circuit-generating notebooks for several of these applications~\cite{coffrin2023quantum}. These notebooks generate quantum circuits that implement Hamiltonian dynamics using pyLIQTR~\cite{Obenland_pyLIQTR} for the quantum signal processing algorithm (QSP) and OpenFermion~\cite{mcclean2020openfermion} for Trotterization. While Ref.~\cite{bartschi2024potential} details several quantum computing applications with scientific utility, we limit our focus to the simulation of Hamiltonians relevant to the search for Kitaev quantum spin liquids (KQSL) at the National High Magnetic Field Laboratory (MAGLAB) at LANL. The particular application benchmark we consider is the dynamical evolution of $\mathrm{\alpha-RuCl_3}$ under a time-varying magnetic field, which is a candidate KQSL material being studied at the MAGLAB (see Chapter 2 of Ref.~\cite{bartschi2024potential} for details). Additionally, we consider the Kitaev model in hexagonal and triangular lattice geometries and the transverse-field Ising model (TFIM) in square, triangular, and cubic lattice geometries. The Kitaev model and TFIM examples provide a range of useful quantum resource comparisons to $\mathrm{\alpha-RuCl_3}$ that enable us to get a sense for resource scaling as a function of system size. 
Specific details about these application instances and quantum algorithm implementations can be found within LANL's open-source repository\footnote{We generated all application circuits using pull request \#47 of the qc-applications repository, which can be found at https://github.com/lanl-ansi/qc-applications/pull/47. The specific notebooks from this repository that were used to generate simulation circuits for the Hamiltonians of interest (see main text) are MagneticLattices.ipynb and RuClExample.ipynb. We used all of the default settings of these notebooks except for the required energy precision in MagneticLattices.ipynb, which we set to $10^{-8}$ in all cases (it was previously inconsistent between Trotter-Suzuki and QSP implementations). On the other hand, we kept the energy precision used in RuClExample.ipynb to its default value, $10^{-3}$, since this value was experimentally motivated.}.

\begin{figure*}[t]
    \centering
    \begin{minipage}{0.49\linewidth}
        \centering
        \includegraphics[width=0.9\linewidth]{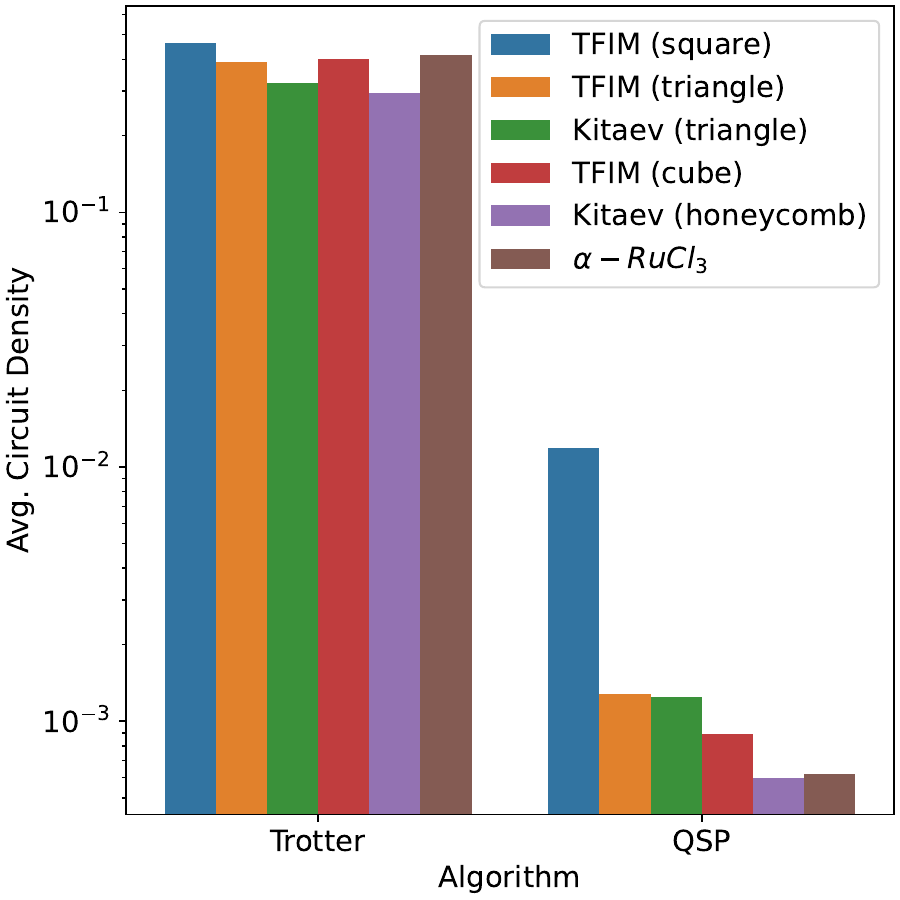}
    \end{minipage}
    \begin{minipage}{0.49\linewidth}
        \centering
        \includegraphics[width=0.9\linewidth]{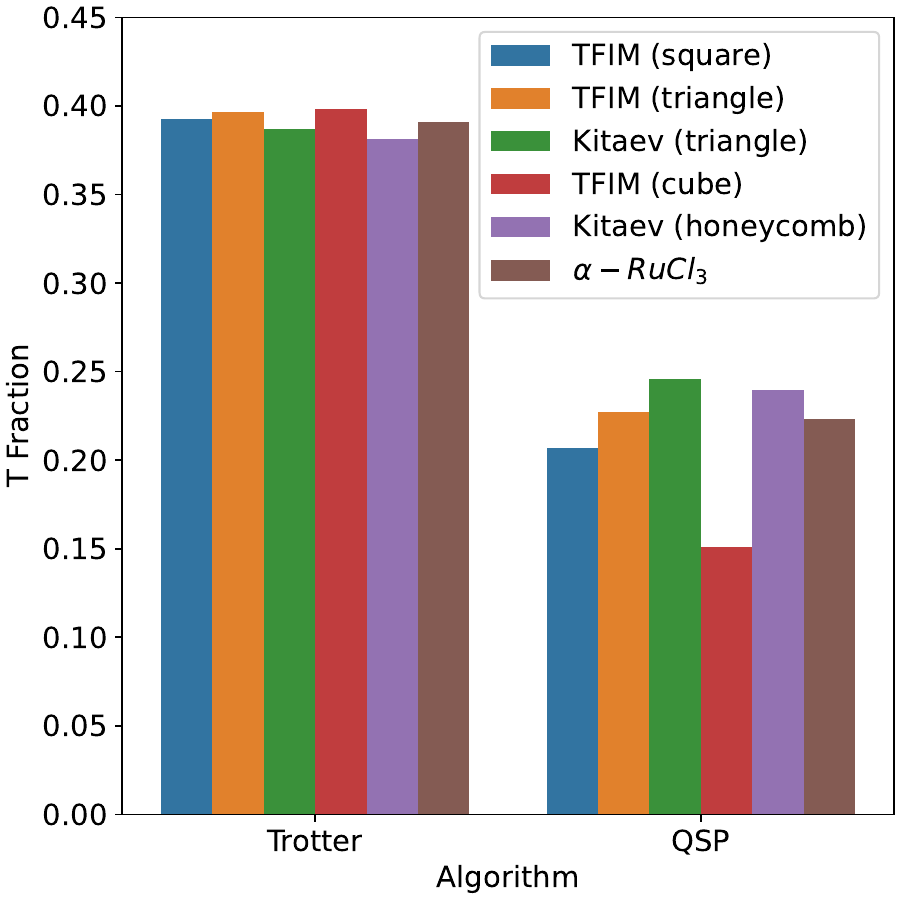}
    \end{minipage}
    \caption{(a) Average circuit densities and (b) and T fractions for the quantum simulation benchmark circuits we consider, organized by algorithm (Trotterization vs. QSP) and in order of increasing numbers of logical qubits (see Table~\ref{tab:lre}).}
    \label{fig:ler_bar}
\end{figure*}

For each application instance, fully decomposed Clifford+T circuits are generated for each key algorithmic sub-unit (subcircuit), e.g. a single Trotter step for second-order Trotter-Suzuki or a block encoding step for QSP. Then, logical-layer resource estimates (LRE) and fault-tolerance-layer resource estimates (FTRE) are obtained for whole circuits by first calculating them independently for each subcircuit and later combining the results. Details on how we calculate resource estimates for whole circuits from those of individual subcircuits can be found in Sec.~\ref{sec:subcircuit}. Table~\ref{tab:lre} contains LRE for the applications under consideration, including the number of repetitions for each key algorithmic sub-unit, the number of logical qubits, the total number of gates, the total number of T gates, and the circuit depth. We remark that, although these gate counts are well beyond the limitations of the Lattice Surgery Compiler (see Sec. 3 of Ref.~\cite{leblond2024realistic} for a scaling analysis of its compilation time per gate for randomly generated circuits), our pipeline is scalable because (as is commonly the case) the circuits are decomposable into a relatively small number of independently compilable sub-circuits that are repeated many times.

In Table~\ref{tab:lre}, we also include two high-level metrics, derived from the LRE, that we aim to suggest should be used by smart compilers to discriminate between applications that would benefit from direct Clifford+T compilation and those that would benefit from SPBC, all before needing to compile. The circuit density is calculated as $\frac{\mathrm{\#\:Gates}}{(\# \:LQ)\times(\mathrm{Depth})}$ (the CNOT gate contributes doubly to the gate count) and the T fraction is the percentage of gates in the circuit which are T gates. Bar charts for both of these derived quantities can be found in Fig.~\ref{fig:ler_bar}. As the main conclusions of this work will be primarily driven by differences in circuit densities between Trotter-Suzuki and QSP implementations, references to Fig.~\ref{fig:ler_bar} will be weaved throughout the ensuing discussion.

\section{Two Approaches to Surface Code Compilation}
\label{sec:twoapproaches}

\subsection{Key Concepts and Terminology}

In the surface code, the fundamental unit of quantum memory is a \textit{tile}, a $d \times d$ array of physical qubits (perhaps with ancillary physical qubits to extract error syndromes) capable of storing one protected logical qubit \cite{litinski2019game}. Here $d$ is the code distance. Tiles are used to store information in logical qubits (in surface code \textit{patches}), move quantum information, and prepare special states (``magic'' states) needed to implement some logical operations. The \textit{spatial footprint} of a computation is the maximum number of simultaneously existing tiles.

Logical circuit operations are implemented in terms of primitive \textit{lattice surgery} operations. A \textit{logical clock cycle} or \textit{time slice}, consisting of $d$ rounds of syndrome extraction, is the time needed to perform a set of simultaneous primitive lattice surgery operations. The \textit{computation time} is $\tau_\text{total} \times d$, where $\tau_\text{total}$ is the number of time slices in the compiled circuit. For a given hardware architecture, the computation time is proportional to the wall-clock run time.

The \textit{space-time footprint} of a computation is the computation time multiplied by the spatial footprint, which is a proxy for the total resources to implement a computation.  A \textit{tile-slice} is the space-time footprint of 1 tile over 1 time slice. A tile is \text{active} during a slice if it is being used to prepare, store, or manipulate a protected quantum state (this includes idling tiles but not uninitialized or measured-out tiles). The \textit{active volume} is the number of active tile-slices in a computation.

\subsection{Sequential Pauli-Based Computation}
\label{sec:spbc}
The first lattice surgery compilation technique that we consider relies upon eliminating Clifford gates from the input logical circuit and results in a circuit consisting only of $\pi/8$ Pauli product rotations (PPRs) and Pauli product measurements (PPM).  These can be implemented using the native multi-body measurement (lattice surgery) instruction set of surface codes together with magic states supplied by an appropriate distillation or cultivation protocol~\cite{fowler2018low,litinski2019game,litinski2019magic,gidney2024magic}. This technique, popularized by Litinski in Ref.~\cite{litinski2019game}, is sometimes referred to as sequential Pauli-based computation (SPBC) because Clifford elimination comes at the cost of effectively sequentializing circuit execution, though refinements have been made to Pauli-based computation that incorporate some degree of logical parallelism~\cite{beverland2022assessing, silva2024multi}. In this work, we do not make use of these refinements but instead capture two extremes by comparing fully sequential SPBC with a highly parallelizable direct Clifford+T compilation scheme (to be described).

We use a standard circuit to implement one of the $\pi/8$ PPRs resulting from Clifford elimination in SPBC, an example of which can be found in Fig. 7 of Ref.~\cite{litinski2019game}. In this circuit, a PPM is first performed between data qubits and a magic state. While this PPM is typically assumed to be accomplishable in one time slice using lattice surgery with an appropriate surface code layout enabling the direct measurement of generic Pauli products (such as the `fast' layout from Ref.~\cite{litinski2019game}), we suspect that the generic PPM instruction set that is popular in surface code compilation theory will be difficult to implement in practice due to the non-triviality of decoding very large patches while dealing with correlated logical errors~\cite{moflic2024constant, skoric2023parallel, cain2024correlated, lin2025spatially}, especially for utility-scale algorithms which may have very large registers. For this reason, in this study we add an overhead of $\tau_{\mathrm{PPM}}$ in the number of time slices to implement PPMs using cat states and Clifford transformations according to the scheme from Ref.~\cite{haner2022space}\footnote{We note that so-called cat states in Ref.~\cite{haner2022space} are more conventionally known as GHZ states.}. Accordingly, a logical cat state can be created in two time slices and a Z-product measurement can be performed in in one time slice using an X-cat state as a resource~\cite{haner2022space}. With this in hand, one can obtain generic PPMs through local Clifford transformations, which require both an $S$ gate and a Hadamard gate in the worst-case scenario. Assuming three time slices for a Hadamard gate~\cite{geher2024error} and two time slices for an S gate~\cite{blunt2024compilation}, this worst-case scenario requires five time slices for local Clifford transformation, resulting in $\tau_{\mathrm{PPM}} = 8$\footnote{An alternative decomposition in a similar spirit can be found in Ref.~\cite{moflic2024constant}.}. The decomposition of PPRs into pair-wise lattice surgery operations as described has the additional benefit of requiring the same surface code instruction set as is required by the direct Clifford+T compilation scheme that we employ (see below), thus providing the best avenue towards an apples-to-apples resource comparison between the two schemes. 

The most sensible surface code layout to implement SPBC using the decomposition we have described is a linear layout with a 1:1 ratio of data qubits to ancilla qubits (nearly identical to Litinski's `intermediate block' from Fig. 13 of Ref.~\cite{litinski2019game} except with different resource states). If we assume that XX interactions occur horizontally on the architecture and ZZ interactions occur vertically, as is the case in Fig. 13 of Ref.~\cite{litinski2019game}, then all of the operations required by the decomposition above can be performed without needing to pay the price for any patch rotations (except for those that occur during logical Hadamard gates, which have already been included in $\tau_{\mathrm{PPM}}$). We include four extra tiles in the layout to accommodate resource states and their interactions with data qubits: one for the preparation of a Y state (which is used to implement the possible Clifford correction required by the aforementioned PPR circuit), one for the storage of a magic state, and two ancillae tiles to mediate interactions between the data qubits and these resource states. Thus, the spatial overhead for the logical block is calculated simply as $n_{\mathrm{logical}} = 2\times (\# \mathrm{LQ} + 2)$. 

Given the simplicity of this scheme, there is no need to explicitly compile SPBC. The computation time can be estimated to good accuracy based on the number of T gates in the circuit and the cost of a worst-case PPM. Our formula for the number of time slices in the logical compilation is $\tau_{\mathrm{logical}} = \tau_{\mathrm{PPM}}\times[1.5 (\mathrm{\#\; T\; gates}) + \mathrm{\#\; LQ}]$, where the factor of 1.5 comes from the approximation that a Clifford $\pi/4$ PPR correction is required 50\% of the time\footnote{In order to avoid real-time transpilation, we assume that this correction is really performed on the architecture and is not eliminated from the circuit.}. The number of distillation factories needed in the computation is correspondingly simple: $N = \lceil\tau_{D}/\tau_{\mathrm{PPM}}\rceil$, where $\tau_{D}$ is the number of time slices required for each factory to produce one magic state. In everything that follows, we assume the same two-level 15-to-1 factory that was used in Ref.~\cite{leblond2024realistic}, which in turn uses the compilation to pairwise lattice surgery instructions for 15-to-1 distillation that was given in Ref.~\cite{beverland2022assessing}\footnote{We do not consider magic state cultivation in this study because the application circuits we consider have huge numbers of T gates that require much lower distilled state error rates than $2\times 10^{-9}$, see Tables~\ref{tab:lre}, \ref{tab:FTRE1} and \ref{tab:FTRE2}.}. See Appendix~\ref{sec:re_clifft} for a summary of the methodology of Ref.~\cite{leblond2024realistic}, including a brief description of factory selection.

Beyond these assumptions, we assume $N$ additional tiles for magic state storage and one warm-up distillation cycle ($\tau_{\mathrm{total}} = \tau_{\mathrm{logical}} + \tau_D$ time slices) to supply magic states for the first set of PPRs, which is consistent with the \textit{min-storage} approach introduced in Ref.~\cite{leblond2024realistic} and described in Appendix~\ref{sec:re_clifft}. The quantification of logical error, as well as the optimization of resources subject to a total error constraint, follows a similar procedure to the one outlined in Appendix~\ref{sec:re_clifft}, except simplified in that we take $\epsilon_{\mathrm{logical}} = n_{\mathrm{logical}} \tau_{\mathrm{logical}} P(d)$ and $\epsilon_{\mathrm{storage}} = N \tau_D (\lceil \tau_{\mathrm{total}} / \tau_D \rceil - 1)P(d)$, where $P(d)$ is the logical error rate for the surface code at distance $d$. In the previous formula, the quantity in parenthesis is the number of distillation cycles present in the whole computation. 

\subsection{Direct Clifford+T Compilation}
\label{sec:direct}
Sequential Pauli-based computation (SPBC) is a popular technique, but since it sequentializes the logical circuit, it is worth comparing its resource performance with schemes that can exploit logical parallelism. Its refinements do admit some degree of fault-tolerance-layer parallelism where PPMs can be simultaneously scheduled~\cite{silva2024multi}, but the PPMs resulting from Clifford elimination in large application circuits are expected to  generically have support over the whole register, precluding them from being simultaneously routed. Therefore, the second technique we consider is the direct Clifford+T compilation approach detailed in Ref.~\cite{leblond2024realistic}, summarized in Appendix~\ref{sec:re_clifft}, and implemented within the Lattice Surgery Compiler~\cite{watkins2023high}. In contrast to SPBC, this scheme is able to translate logical-layer parallelism into fault-tolerance-layer parallelism at the expense of needing to explicitly perform all Clifford operations\footnote{We note that Ref.~\cite{hirano2025locality} recently introduced a scheme that performs some Clifford elimination while retaining locality and thus fault-tolerance-layer parallelizability, thus combining the best of both worlds. We leave resource benchmarking for this scheme to future work.}.

The direct Clifford+T methodology we use in this study is very similar to the one used by Ref.~\cite{leblond2024realistic} except for minor modifications. While we previously used the catalytic S gate teleportation circuit from Ref.~\cite{fowler2012surface}, we presently assume a protocol based on Gidney's cheap twist-based Y state initialization technique~\cite{gidney2024inplace, blunt2024compilation}. This protocol not only has a substantially lower overhead than the catalytic one but also enables a more direct comparison to the variant of the SPBC protocol that we employ, which also relies on local Clifford transformations but cannot use the catalytic S gate circuit due to its need to perform many parallel S gates in a linear architecture. Also, instead of using the 1-lane layout (specified in Sec. 2 of Ref.~\cite{leblond2024realistic}) we focus on the 1-lane-condensed layout in Sec.~\ref{sec:results}, although results were comparable for 1-lane (see Table~\ref{tab:FTRE3} for fault-tolerance-layer resource estimates that use the 1-lane layout instead of 1-lane-condensed). Lastly, the tiles that were originally reserved for Y states in our layouts now serve as additional tiles that are reserved for magic states.

Since magic state consumption rates are not constant throughout individual circuits in the direct Clifford+T compilation scheme, we use the \textit{min-storage} approach to distillation-storage resource optimization introduced in Ref.~\cite{leblond2024realistic}. In brief, this strategy minimizes the resources needed for magic state storage by allocating enough magic state factories to provide the maximum number of magic states consumed in any one distillation cycle, and temporarily ``turning off" some of the factories for cycles that consume fewer magic states. Despite the fact that this technique requires a greater spatial footprint for magic state distillation, the results of Ref.~\cite{leblond2024realistic} showed that it admits of orders-of-magnitude lower space-time footprints for circuits with fluctuating magic state consumption rates than do approaches that hold the number of factories constant during computation. For further details, see Appendix~\ref{sec:re_clifft}.

\subsection{Fault-Tolerance-Layer Resource Estimates}
\label{sec:ftre}
In this study, we consider the following high-level fault-tolerance-layer resource estimates (FTRE): $\tau_{\mathrm{total}}$ represents the total number of time slices (logical clock cycles) required for complete fault-tolerant circuit execution, $n_{\mathrm{total}}$ represents the total number of surface code tiles (including those utilized by distillation factories and for magic state storage), and $d$ is the code distance required to execute the circuit beneath a total error budget of $\epsilon_\mathrm{tot} = 0.01$. As we consider two-level magic state distillation, we specify code distances $d_1$ and $d_2$ for the first and second level, respectively; the code distance utilized in logical computation is $d = d_2$. We refer to the number of magic state factories as $N$ and the logical error rate for distilled magic states as $P_T$. We refer to the active volume of computation as $V$. We divide $\epsilon_\mathrm{tot}$ into contributions from computation ($\epsilon_{\mathrm{logical}}$), magic state production ($\epsilon_{\mathrm{dist}}$), and magic state storage ($\epsilon_{\mathrm{storage}}$). See Appendix~\ref{sec:re_clifft} for a review of how resource estimates for individual circuits are derived from the outputs of direct Clifford+T lattice surgery compilation according to the methodology introduced in Ref.~\cite{leblond2024realistic}. In the next section, we will describe how this framework is generalized to the setting where a circuit is composed of several repeating subcircuits.

\section{Resource Analysis for Subcircuits}
\label{sec:subcircuit}
Since the resource estimates we provide for SPBC depend only on the number of logical qubits and the number of T gates required by a given circuit, there is no subtlety involved in tallying resource costs subcircuit-by-subcircuit to yield resource costs for an entire algorithm. 
This is not the case, however, in the direct Clifford+T compilation scheme.  Since lattice surgery compilation needs to be performed separately for each subcircuit due to scaling limitations of the Lattice Surgery Compiler (see Sec. 3 of Ref.~\cite{leblond2024realistic}), it is necessary to combine the FTRE of separately compiled subcircuits. To that end we generalize the protocol described in Ref.~\cite{leblond2024realistic} for resource estimation in direct Clifford+T compilation, summarized in Appendix~\ref{sec:re_clifft}, to the setting where circuits are composed of several subcircuits executing serially on the architecture\footnote{This is an assumption that we have inherited from the tools that we employ to compile quantum simulation circuits (see Sec.~\ref{sec:app}).}. Therefore, computation time is additive: $\tau_{\mathrm{total}} = \sum_i n_{i} \tau_i$, where the sum is over subcircuits and $n_i$ is the number of occurrences for subcircuit $i$ present in the whole circuit. The active volume and logical error rates are similarly additive ($V_{\mathrm{total}} = \sum_i n_{i} V_i$ and $\epsilon_{\mathrm{total}} = \sum_i n_{i} \epsilon_i$). Lastly, the number of logical tiles required by the whole algorithm is its maximum over subcircuits, including contributions from the logical layout, magic state distillation, and magic state storage.

In the direct Clifford+T scheme, for the sake of simplicity in resource optimization, we require the same code distances to be used for all subcircuits\footnote{A consequence of this requirement is that the specification for a single magic state factory is uniform throughout the whole algorithm.}.
On the other hand, we do not require data qubits to remain in a consistent logical layout between subcircuit executions. The layout family that we use (see Sec.~\ref{sec:twoapproaches}) is parameterized by the number of data qubits targeted by each subcircuit, and we expect that allowing the layout to shrink for subcircuits operating on sub-registers comes with active volume savings (which, in turn, means lower logical error rates), although we did not explicitly verify this. We also do not assume the same amount of space dedicated to magic state distillation and storage for all subcircuits, as they may have different magic state consumption requirements\footnote{The number of magic state distillation and storage tiles \textit{in use} varies not only between subcircuits but also within the execution of individual subcircuits according to the \textit{min-storage} approach introduced in Ref.~\cite{leblond2024realistic} and described in Appendix~\ref{sec:re_clifft}. However, the number of tiles \textit{allocated} to distillation and storage for each subcircuit is constant, and determined by the maximum-consumption distillation cycle of each subcircuit. Therefore, magic state factories and storage tiles are dynamically allocated on a subcircuit-by-subcircuit basis, but are static within subcircuits.}. Note, however, that the number of magic state factories $N$ reported for each (Application, Algorithm) pair in Table~\ref{tab:FTRE2} and~\ref{tab:FTRE3} is its maximum over the whole algorithm's execution.

\begin{table*}[t]
\centering
\caption{For each application and for each metric of interest, we show the combination of quantum algorithm (Trotter or QSP) and compilation scheme (SPBC or direct Clifford+T) that produces a minimal value for that metric. `min-storage' designates the distillation-storage resource optimization approach from Ref.~\cite{leblond2024realistic} that we have chosen to use in this study.}
\begin{tabular}{|c|c|c|}
\toprule
 Application & Computation Time ($\tau_{\mathrm{total}} \times d$) & Space-Time Footprint ($n_{\mathrm{total}} \times \tau_{\mathrm{total}} \times d^3$)\\
\midrule
TFIM (square) & (QSP, Clifford+T (min-storage)) & (QSP, SPBC) \\
TFIM (triangle) & (Trotter, Clifford+T (min-storage)) & (QSP, SPBC) \\
TFIM (cube) & (Trotter, Clifford+T (min-storage)) & (QSP, SPBC) \\
Kitaev (triangle) & (QSP, Clifford+T (min-storage)) & (QSP, SPBC) \\
Kitaev (honeycomb) & (QSP, Clifford+T (min-storage)) & (QSP, SPBC) \\
$\mathrm{\alpha-RuCl_3}$ & (Trotter, Clifford+T (min-storage)) & (Trotter, Clifford+T (min-storage)) \\
\bottomrule
\end{tabular} 
\label{tab:optimal}
\end{table*}

Because the logical layout may change between subcircuits according to the number of logical qubits targeted by each one, we include an active volume contribution for idling data qubits that are not being targeted by a given subcircuit and are therefore not present in its logical layout specification. Further, while we include zero overhead for shuffling layouts in between subcircuit executions, we do include one warm-up distillation cycle in between subcircuit executions as well as a contribution to active volume from idling data qubits during this cycle. We expect that layout shuffling can occur during this time with minimal extra cost. We also assume zero overhead for transporting magic states to their designated locations in the logical block during subcircuit execution. However, we do include the spatial and active volume requirements for magic state storage (the states produced in the $(n-1)$th cycle are stored outside the layout until their consumption during the $n$th cycle), which we expect to approximate the magic state transportation cost.

\section{Results}
\label{sec:results}

The primary question of interest in this study is whether, for certain practical application benchmarks of interest, the benefit of eliminating Clifford gates from the logical circuit outweighs the reduced parallelism of SPBC. In this section, we compare fault-tolerance-layer resource estimates (FTRE) for quantum simulation tasks derived from representatives of two leading approaches to surface code lattice surgery compilation: direct Clifford+T compilation (`direct') and Pauli-based computation. Details of the assumptions we make in either case were fully described in Section~\ref{sec:twoapproaches}. Preliminarily, we note that the results of Ref.~\cite{leblond2024realistic} suggested that the `direct' scheme may have an asymptotic benefit in computation time compared to SPBC for a particular series of Trotter-Suzuki ground state estimation (GSE) circuits~\cite{kornellsome}. In that work it was found that in the `direct' scheme the number of time slices per circuit T gate decreased as the GSE circuits grew larger (i.e., as the system size for the corresponding application increased), whereas for SPBC the number of time slices per T gate is constant. Nevertheless, the impact of direct compilation on total space-time resources was not obvious due to the much higher cost of magic state distillation (needed to accomodate many parallel T gates in direct Clifford+T compilation), exacerbated by the variation in magic state consumption rate that we find for real circuits\footnote{See Sec. 5 of Ref.~\cite{leblond2024realistic} for a detailed discussion of the implications of this variation, and Appendix~\ref{sec:re_clifft} for a summary.}. Also, while the benefits of the `direct' scheme are intuitive for circuits having a high degree of logical parallelizability, it is expected that SPBC will yield the better performance for logical circuits that have little parallelizability because in this regime the benefit of Clifford elimination should dominate. Because of this trade space complexity, it is important to know concretely which scheme performs better for specific benchmark applications, and to what extent this depends upon the choice of quantum algorithm (e.g. QSP, which is fairly serial, or Trotter, which is highly parallelizable; see Fig.~\ref{fig:ler_bar}a for average circuit densities in either case). 

As a summary of our results, Table~\ref{tab:optimal} shows the optimal (algorithm, scheme) combination for each application benchmark under two different metrics of interest: the computation time ($\tau_{\mathrm{total}} \times d$) and the space-time footprint ($n_{\mathrm{total}} \times \tau_{\mathrm{total}} \times d^3$). According to these results, QSP with SPBC yielded the lowest space-time footprints for all applications except for $\mathrm{\alpha-RuCl_3}$, for which Trotterization with the `direct' scheme was found to be optimal. On the other hand, the `direct' scheme yielded the lower computation time for all applications, although the optimal choice of quantum algorithm in this case appears to be application-dependent. While these high-level results are interesting on their own, a more detailed breakdown will allow us to gain greater insight\footnote{Tables~\ref{tab:FTRE1} and~\ref{tab:FTRE2} in Appendix~\ref{sec:tables} contain the full fault-tolerant resource estimates (FTRE) under discussion in this section}.

\begin{figure*}[htbp]
    \centering
    \includegraphics[width=\textwidth]{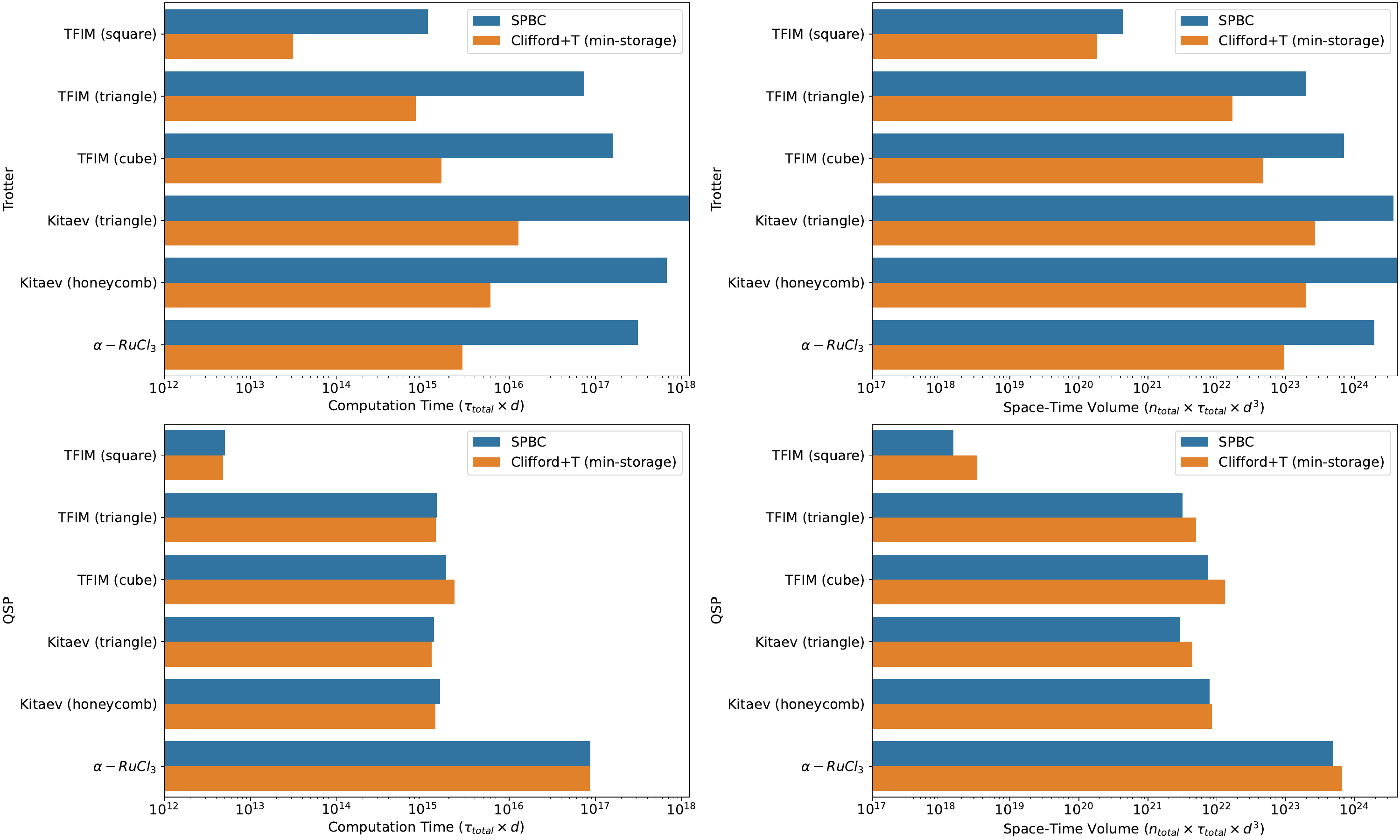}
    \captionsetup{skip=10pt}
    \caption{Computation time ($\tau_{\mathrm{total}} \times d$, left column) and space-time footprint ($n_{\mathrm{total}} \times \tau_{\mathrm{total}} \times d^3$, right column) for Trotterization (top row) and quantum signal processing (bottom row) implementations of quantum simulation for $\mathrm{\alpha-RuCl_3}$ as well as the transverse-field Ising model and Kitaev models in several geometries. As discussed in the caption of Table~\ref{tab:optimal}, `min-storage' designates the distillation-storage resource optimization approach from Ref.~\cite{leblond2024realistic} that we have chosen to use in this study.}
    \label{fig:barchart}
\end{figure*}

\subsection{Computation Times}
In Fig.~\ref{fig:barchart}, we show the computation times (left column) and space-time footprints (right column) for Trotter-Suzuki (top row) and QSP (bottom row) implementations of each application benchmark. The top-left panel demonstrates one of the main insights of this study, namely, that the `direct' scheme yields a roughly 100x reduction in computation time compared to SPBC for the Trotter implementations\footnote{Computation times in the `direct' scheme are even lower when using the 1-lane surface code layout than 1-lane-condensed; see Table~\ref{tab:FTRE3} for details.}. Perhaps interestingly, we see that the reduction is a bit less pronounced for TFIM (square), which involves a smaller number of logical qubits than the other applications and yet has similar circuit density (see Fig.~\ref{fig:ler_bar}). This observation is consistent with the expectation that each layer of logical gates in dense circuits translates to a number of time slices that is $O(\sqrt{x})$ in the number of logical qubits (see Sec. 3 of Ref.~\cite{leblond2024realistic} for results on randomly generated circuits). We confirm this expectation for our circuits in Fig.~\ref{fig:powerlaw}a, which shows that the number of time slices per logical layer increases slightly faster than this expectation. The exponent for the power law fit shown in Fig. 4a is similar to those found in Fig. 5a of Ref.~\cite{leblond2024realistic}, suggesting a similarity of character between these Trotter-Suzuki circuits and those randomly generated circuits, at least as far as direct Clifford+T to lattice surgery compilation is concerned. Accordingly, we expect the separation between computation times resulting from the `direct' scheme and SPBC to increase with problem size for Trotter-Suzuki circuits. 

This dramatic reduction in computation time for Trotter circuits is in sharp contrast to what we find for QSP circuits in the bottom-left panel of Fig.~\ref{fig:barchart}. There, differences in computation time appear nearly insignificant between direct Clifford+T and SPBC implementations of QSP for these applications, although they do appear to loosely correlate with differences in the T fraction between circuits [see Fig.~\ref{fig:ler_bar}(b) and Table~\ref{tab:lre}]. For instance, Fig.~\ref{fig:ler_bar}(b) shows a relatively low T fraction for TFIM in the cubic geometry, and this circuit appears to demonstrate the greatest relative benefit of computation time from SPBC over direct Clifford+T. The converse is almost true for the circuits corresponding to the Kitaev model in triangular and honeycomb lattices, although the order is swapped, showing that for these circuits T fraction alone does not indicate which scheme yields lower computation times. That said, these differences are very minor, especially as compared with the ones we observed for Trotter-Suzuki circuits.

\begin{figure*}
    \centering
    \begin{minipage}{0.49\linewidth}
        \centering
        \includegraphics[width=0.9\linewidth]{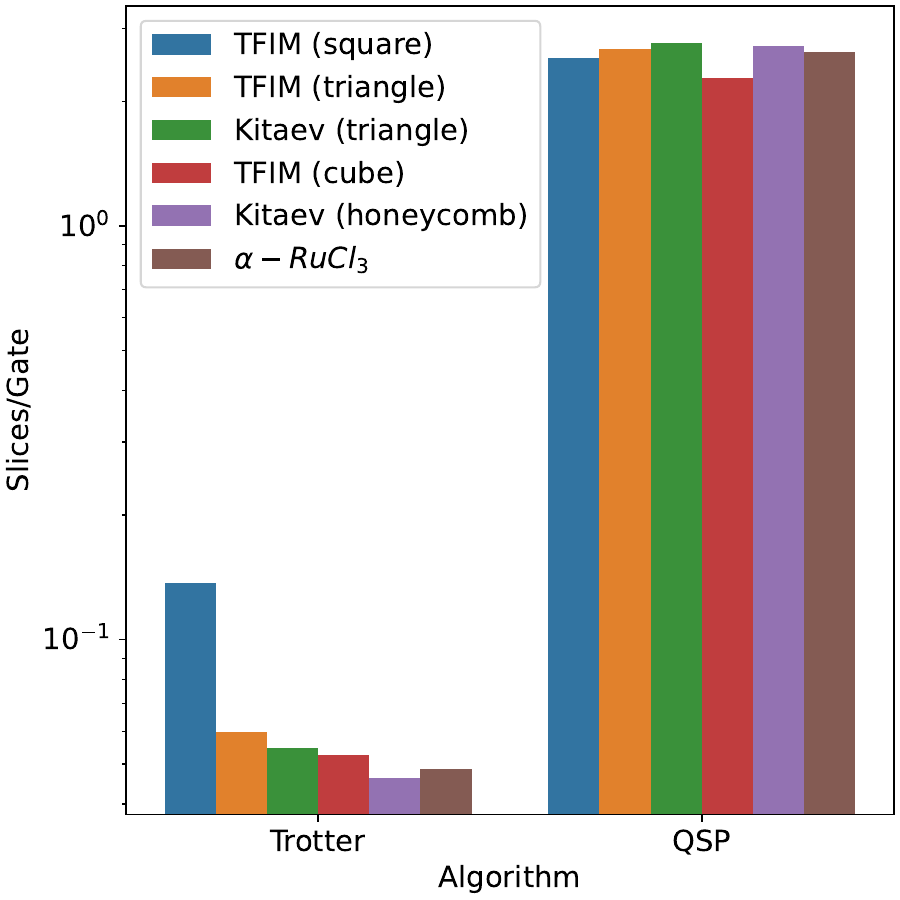}
    \end{minipage}
    \begin{minipage}{0.49\linewidth}
        \centering
        \includegraphics[width=0.9\linewidth]{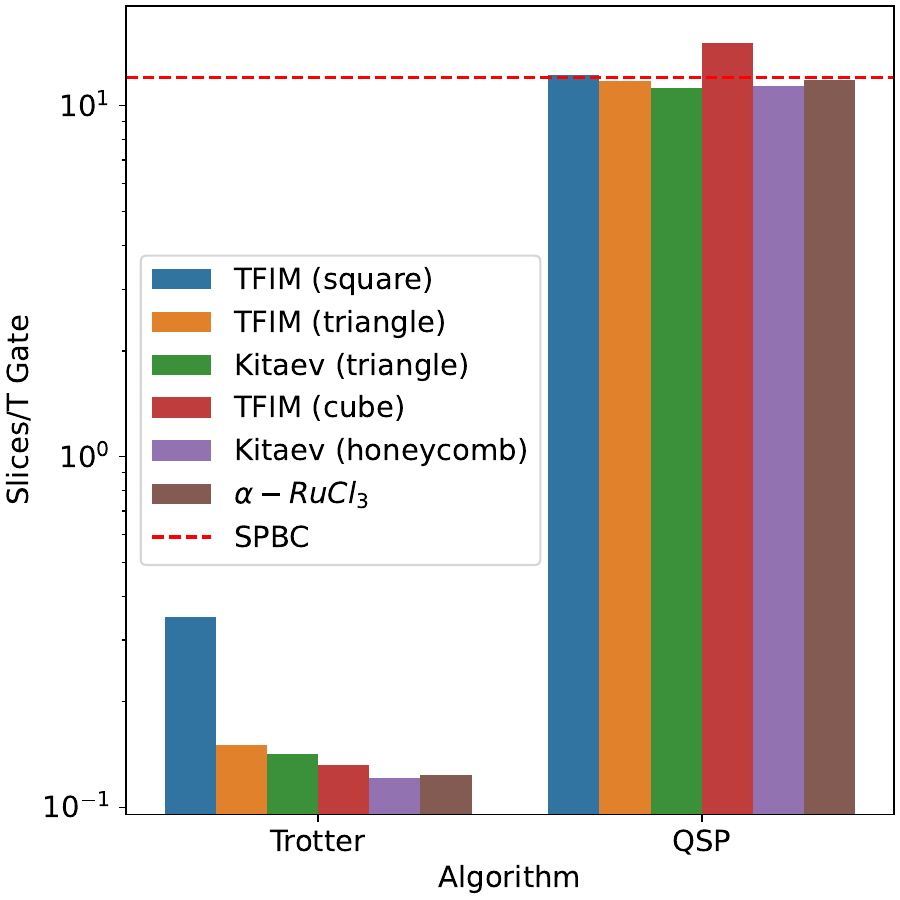}
    \end{minipage}
    \caption{(a) The average number of lattice surgery time slices per gate from the logical (input) circuit. (b) The total number of time slices for the compiled circuit divided by the number of T gates in the input circuit, for direct Clifford+T compilation of the applications of quantum simulation considered in this study.}
    \label{fig:slices}
\end{figure*}

\begin{figure}
    \centering
    \begin{subfigure}{\linewidth}
        \centering
        \includegraphics[width=\linewidth]{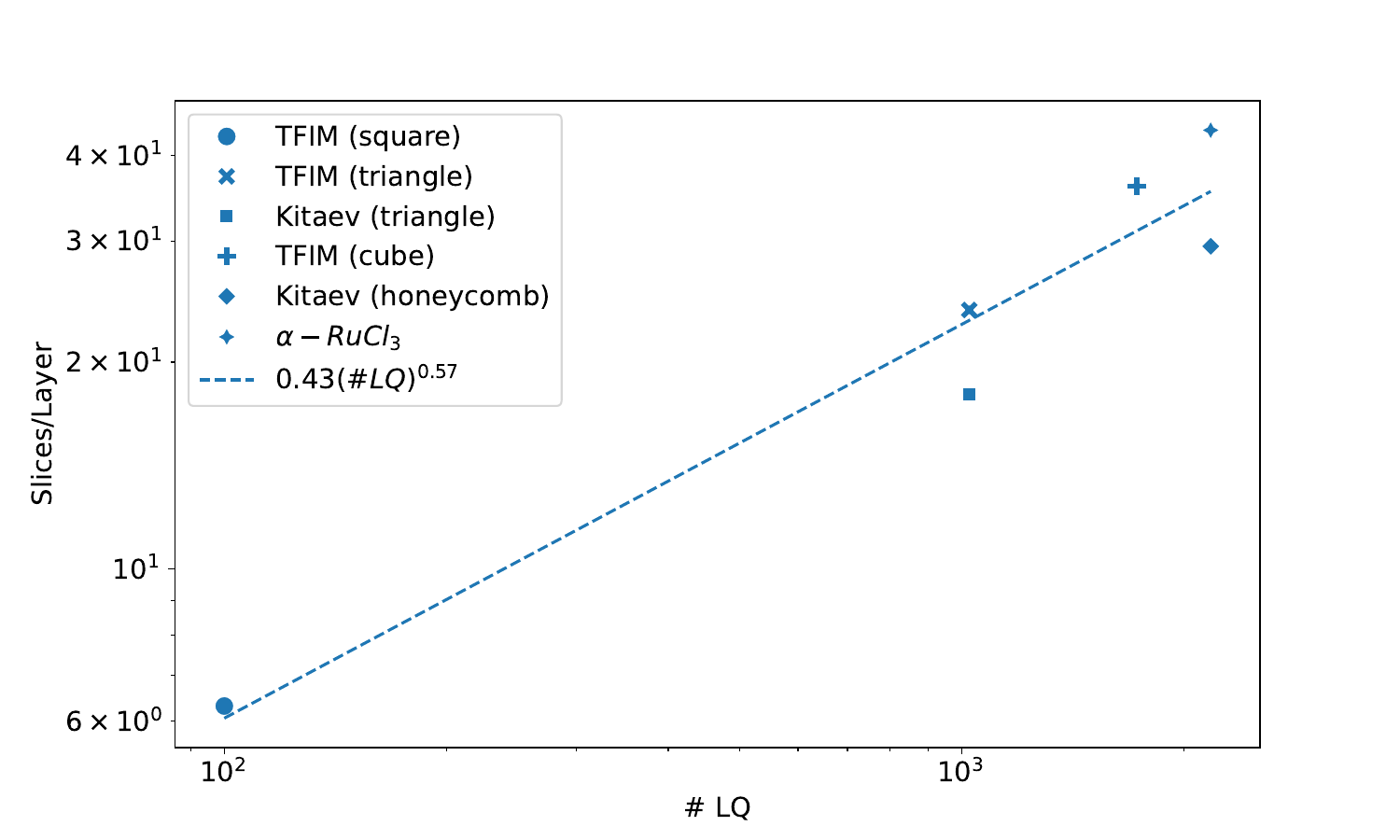}
        \label{fig:powerlaw1}
    \end{subfigure}
    \begin{subfigure}{\linewidth}
        \centering
        \includegraphics[width=\linewidth]{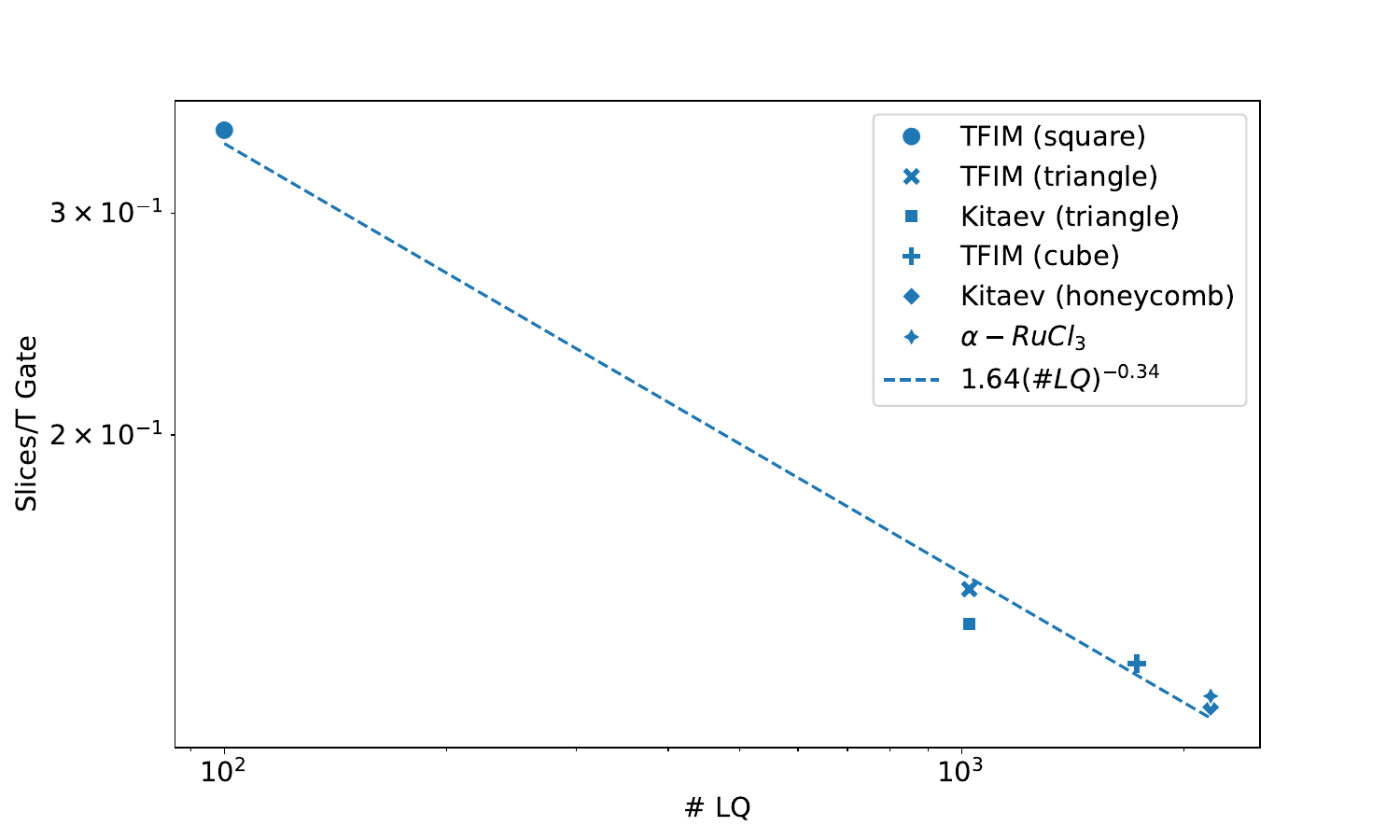}
        \label{fig:powerlaw2}
    \end{subfigure}
    \begin{subfigure}{\linewidth}
        \centering
        \includegraphics[width=\linewidth]{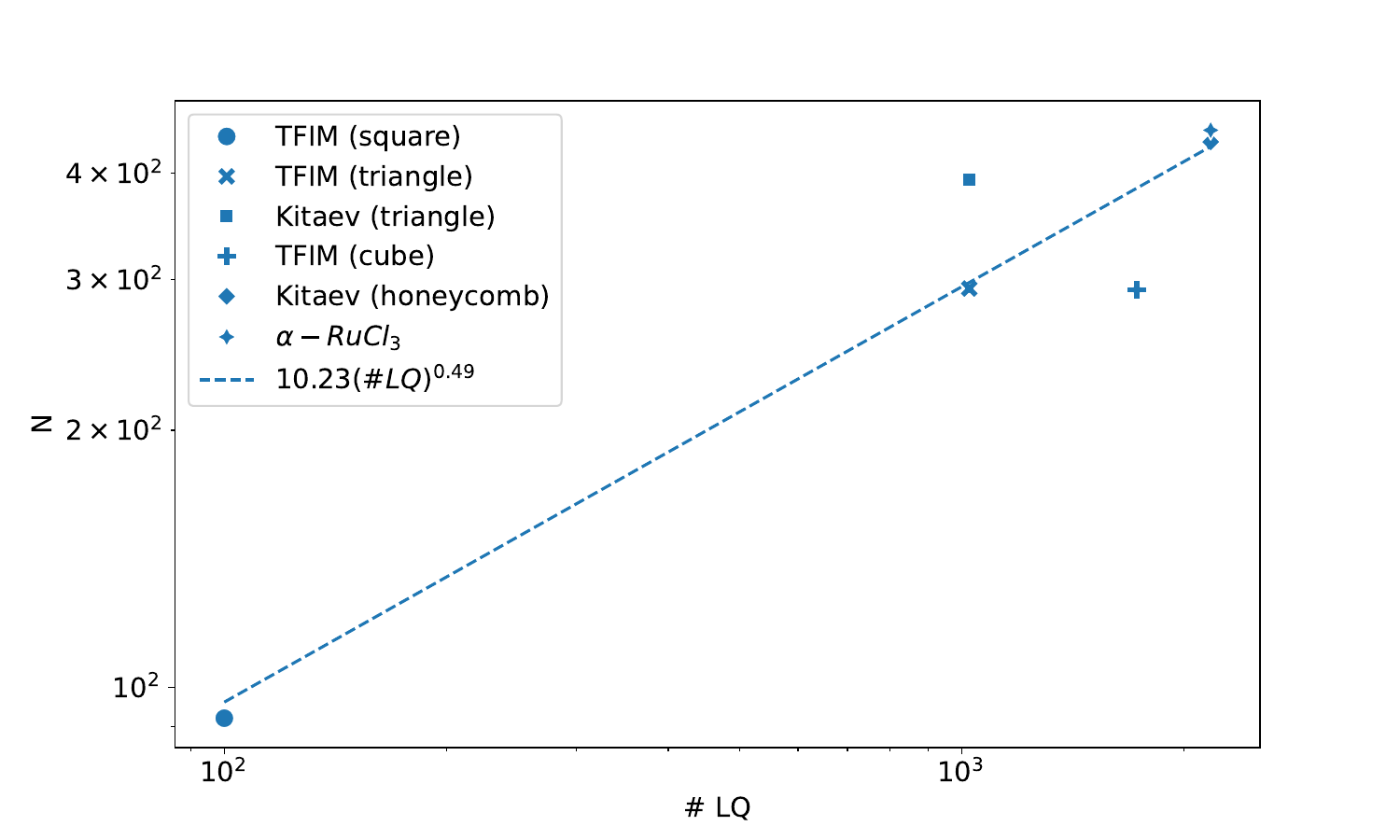}
        \label{fig:powerlaw3}
    \end{subfigure}

    \caption{For the direct Clifford+T compilation of the Trotter-Suzuki implementation of each application benchmark considered in this study, we show (a) the number of lattice surgery time slices divided by the depth of the input logical circuit, (b) the number of time slices divided by the number of T gates in the logical circuit, and (c) the total number of magic state factories provided by the \textit{min-storage} resource optimization approach, all plotted against the number of logical qubits (\# LQ) required by the circuit (see Table~\ref{tab:lre}). Power law fits to the data are provided in all three cases.}
    \label{fig:powerlaw}
\end{figure}

Accordingly, and as expected, the massive differences in computation time derived from `direct' and SPBC analyses for these Trotter-Suzuki circuits can be explained through their great degree of logical parallelizability as compared with QSP circuits. This is evidenced by their high, non-decreasing average circuit densities [see Fig.~\ref{fig:ler_bar}(a) and Table~\ref{tab:lre}]. While the Trotter-Suzuki circuits we consider have average circuit densities ranging between 29.5\% [Kitaev (honeycomb)] and 46.3\% [TFIM (square)], those of the QSP circuits are much lower and decrease with system size. In fact, the QSP circuits' densities are not much higher than if there were a single gate per layer, underlining the nearly serial character of these circuits. Correspondingly, the average number of time slices \textit{per gate} in the Trotter-Suzuki circuits is small and decreases as the problem size increases, while this metric takes a much higher (and nearly constant) value for QSP (see Fig.~\ref{fig:slices}a). Strikingly, for the larger applications we consider, there are less than 5\% as many time slices as there are gates in the input logical circuit. Thus, it is obvious that, even if SPBC were implemented by directly measuring generic Pauli products in such a way that $\tau_{\mathrm{PPM}} = 1$ (as is often assumed in the literature), direct Clifford+T compilation would yield superior computation times for these circuits (especially since these circuits have T fractions much higher than $\sim 5\%$).

To drive this point home, because the total number of time slices for SPBC is directly proportional to the number of T gates in the input logical circuit, it is appropriate to compare the number of time slices per circuit T gate between QSP and Trotter-Suzuki circuits (Fig.~\ref{fig:slices}b). Because we have conservatively set $\tau_{\mathrm{PPM}} = 8$ and because of the Clifford correction which occurs $\approx 50\%$ of the time, the number of time slices per T gate for SPBC is approximately 12. Thus, if the direct Clifford+T approach yields less than 12 time slices per circuit T gate, it will yield lower computation times than SPBC. Fig.~\ref{fig:slices}b demonstrates that this quantity lies well below the SPBC threshold of 12 for our Trotter circuits, while it lies approximately at the threshold for our QSP circuits. A power law fit to this quantity is shown in Fig.~\ref{fig:powerlaw}b for Trotter-Suzuki circuits, clearly demonstrating that the improvement of `direct' over SPBC is increasing as the number of logical qubits increases.

\subsection{Space-Time Footprints}
When the spatial foot-print is factored in, we find that the relative advantage of direct Clifford+T compilation over SPBC decreases. This is expected not only because the layouts are larger in the `direct' scheme than in SPBC, but also because of the greater magic state production requirements in the former scheme than the latter. Still, the top-right panel of Fig.~\ref{fig:barchart} shows that the space-time footprints of the Trotter circuits are about 10-20x lower in the direct Clifford+T scheme than they are in the SPBC scheme. As before, there is less of a difference for TFIM (square) than for the other application benchmarks. Surprisingly, the space-time footprints are smaller for almost all Trotter circuits when the 1-lane layout is used than when the 1-lane-condensed layout is used (compare Tables~\ref{tab:FTRE2} and~\ref{tab:FTRE3} in Appendix~\ref{sec:tables}). This is counter-intuitive since 1-lane is the larger layout, having a 3:1 ancilla:data tile ratio in the bulk instead of the 5:4 ancilla:data tile ratio of 1-lane-condensed. However, the greater degree of fault-tolerance-layer parallelism enabled by 1-lane over 1-lane-condensed more than makes up for its larger spatial footprint and greater magic state production requirements (the latter of which is reflected in the number of factories, $N$, that was determined by \textit{min-storage} in each case). Very briefly, we also mention that the bottom-right panel of Fig.~\ref{fig:barchart} shows that, for QSP implementations, SPBC yields the lower space-time footprint in all cases. It is perhaps more surprising that the results are competitive than it is to find that SPBC performs better.

Finally, we wish to highlight a couple of insights from Tables~\ref{tab:FTRE1} and~\ref{tab:FTRE2} that help us understand how a $\sim$100x improvement in computation time for these Trotter circuits the `direct' scheme over SPBC becomes a $\sim$10-20x improvement in space-time footprint. First of all, we notice an immensely greater number of factories $N$ in the `direct' scheme than in SPBC (for the `direct' scheme with 1-lane-condensed, $N$ ranges from 92 for TFIM (square) to 449 for $\mathrm{\alpha-RuCl_3}$, while in SPBC $N = 3$). This is due to the fact that, in the former scheme (with \textit{min-storage}), $N$ is equal to the maximum number of magic states consumed within any distillation cycle worth of computation time, while in the latter scheme $N$ is simply calculated as $N = \lceil\tau_{D}/\tau_{\mathrm{PPM}}\rceil$. As can be seen in Fig.~\ref{fig:powerlaw}c, the number of factories determined by \textit{min-storage} for these circuits follows the expected square root law in the number of logical qubits due to the layout requirement that magic states enter from the boundaries and are not produced in the bulk. The fact that the spatial contribution from factories follows a square root law while the layout contribution is linear causes us to expect that the relative contribution from factories becomes small for very large applications. For $\mathrm{\alpha-RuCl_3}$, however, we find that 449 factories still have a substantial impact on the spatial footprint since $n_{\mathrm{total}}$ is approximately 7 times larger in the `direct' scheme with 1-lane-condensed than in SPBC. 

With this accounted for, we find that the `direct' scheme has a 14 times lower number of tile-slices than SPBC. Then, to complete the observed difference in space-time volume, we notice that the required code distances in SPBC and direct Clifford+T are 37 and 33, respectively, owing to the much greater logical computation time (and therefore logical active volume, one of the leading contributors to logical error) in the former than in the latter. The increased code distance adds insult-to-injury to SPBC, causing the 14x lower number of tile-slices in the `direct' scheme to become a 20x lower space-time footprint.  

\section{Conclusion}
\label{sec:conclusion}
In this study, we have presented a quantum resource comparison between a direct Clifford+T to lattice surgery compilation scheme and standard sequential Pauli-based computation (SPBC) for various quantum simulation circuits that have been deemed to have scientific utility. The choices of lattice surgery compilation scheme were meant to capture two extremes: on one hand, we use a compiler to estimate a realistic parallel implementation of Clifford+T circuits on a surface code processor, while on the other hand we perform a simpler Litinski-style analysis that has been adapted to maintain a common surface code instruction set with the Clifford+T scheme. Our major findings are as follows: where the Trotter-Suzuki algorithm is used to implement quantum simulation, direct Clifford+T compilation provides superior computation time \textit{and} space-time footprint, with about two orders of magnitude lower computation time and one order of magnitude lower footprint in most cases. On the other hand, where quantum signal processing (QSP) is used to implement quantum simulation, SPBC is the (slightly)-preferred compilation scheme by space-time footprint, while computation time is very similar between the two lattice surgery compilation schemes. Altogether, we find that the QSP algorithm with SPBC leads to lower space-time footprints than second-order Trotter-Suzuki for most applications while Trotterization with direct Clifford+T compilation leads to the lower footprint for $\mathrm{\alpha-RuCl_3}$, which is the largest application we consider. 
That said, we expect that Trotterization with direct Clifford+T compilation will yield a lower footprint for larger applications than those considered in this study. This expectation is based on two factors: first, QSP implementations show a constant number of total time slices per circuit T gate in the `direct' scheme while this ratio decreases for Trotter-Suzuki implementations. Second, although the spatial contribution to magic state distillation in the `direct' scheme is substantial, it is sub-leading.

It may be said that, wherever we have seen the direct Clifford+T scheme deliver massive benefits over SPBC, we should be tempered by the understanding that this gap could be narrowed if a compiler is used to estimate more realistic costs in the latter scheme by using e.g. the forest packing algorithm~\cite{silva2024multi} to schedule logically parallel PPMs where possible given architectural constraints. While, judging by the results of Ref.~\cite{silva2024multi}, it does seem that some logical parallelism can be captured in the fault-tolerance layer in SPBC for the (relatively) small circuits they consider, it is doubtful that very much parallelization of PPMs is possible for large circuits such as those under consideration in this study. This is primarily due to the fact that entangling gates generally increase the support of PPRs upon commutation. Eventually, the support of PPRs practically extends over the whole register, making opportunities for parallelism very limited. That said, future work should explicitly verify this claim. Additionally, further compiler optimizations could have been applied in the direct Clifford+T compilation scheme~\cite{beverland2022surface, hamada2024efficient}. For example, we expect edge-disjoint path compilation to offer meaningful benefits over the vertex-disjoint path compilation approach utilized here, as our preliminary tests have shown power-law depth improvements for the same series of randomly generated circuits that we found offer similar lattice surgery performance scaling to the Trotter circuits we have considered in this study\footnote{See pull request \# 121 of the Lattice Surgery Compiler, https://github.com/latticesurgery-com/liblsqecc/pull/121, for details.}. That said, and especially since further compiler optimizations can reduce the number of PPR that need to be compiled to the architecture in PBC~\cite{litinski2019game} and since the number of lattice surgery operations required post-decomposition to pair-wise instructions can possibly also be reduced~\cite{moflic2024constant}, future work should implement advanced SPBC techniques into a resource estimation software workflow so that the trade-space between locality and Clifford elimination can be more concretely understood.

Although the spin dynamics applications under consideration in the present study are important due to their understood scientific utility~\cite{coffrin2023quantum}, it is \textit{a priori} unclear whether our qualitative insights apply to quantum dynamics for fermionic Hamiltonians. Although Trotter circuits for fermionic systems are expected to have lower parallelizability than those of spin systems due to the need for mappings from fermions to spins, recent advances such as the use of fermionic SWAP networks to localize interactions at the expense of additional operations may ameliorate this problem~\cite{akahoshi2025compilation}. As such, these circuits may lie between the two extremes studied in the present work. It would be interesting for future work to compare resource costs of fermionic dynamics simulations between direct Clifford+T and SPBC compilation schemes.

Lastly, we acknowledge that the FTRE reported in this study are huge. Though these resources are abstracted from the physical layer, it is straightforward to obtain physical-layer estimates by e.g. multiplying $\tau_{\mathrm{total}}\times d$ by the time (in seconds) for one round of syndrome extraction. The emphasis of this study was to compare resource estimates from two competing approaches to surface code lattice surgery compilation and their implications on FTRE, not to propose expected resource estimates for executing the applications we consider. Our results highlight the need for (a) improved general-purpose surface code lattice surgery compilation techniques and open-source implementations and (b) improved resource optimization techniques for surface code lattice surgery architectures. Our results also perhaps strengthen the case that surface code lattice surgery has too large an overhead for utility-scale quantum computation.

\section*{Acknowledgements}
We acknowledge helpful discussions with Zachary Morrell and Jonhas Colina as we made use of LANL's Quantum Applications Specifications and Benchmarks tool. 

This work was performed as part of the Defense Advanced Research Projects Agency Underexplored Systems for Utility-Scale Quantum Computing (US2QC) program. The manuscript was completed with support from the U.S. Department of Energy, Office of Science, National Quantum Information Science Research Centers, Quantum Science Center. This research used resources of the Compute and Data Environment
for Science (CADES) at the Oak Ridge National Laboratory, which is supported by the Office of Science of the
U.S. Department of Energy under Contract No. DE-AC05-00OR22725.

This manuscript has been authored by UT-Battelle, LLC under Contract No. DE-AC05-00OR22725 with the U.S. Department of Energy. The publisher, by accepting the article for publication, acknowledges that the U.S. Government retains a non-exclusive, paid up, irrevocable, world-wide license to publish or reproduce the published form of the manuscript, or allow others to do so, for U.S. Government purposes. The DOE will provide public access to these results in accordance with the DOE Public Access Plan (http://energy.gov/downloads/doe-public-access-plan). 

\section*{Author Contributions}
T. L. conceived the study, developed the compiler and resource estimation methodology, performed the numerical experiments, and prepared the manuscript. R. B. supervised the project, contributed to discussions and edited the manuscript. Both authors reviewed and approved the final manuscript. Generative AI tools were used to assist with code development. All generated code was reviewed, modified as needed, and validated by the authors.

\bibliographystyle{quantum}
\bibliography{refs}

@article{litinski2019game,
  title={A game of surface codes: Large-scale quantum computing with lattice surgery},
  author={Litinski, Daniel},
  journal={Quantum},
  volume={3},
  pages={128},
  year={2019},
  publisher={Verein zur F{\"o}rderung des Open Access Publizierens in den Quantenwissenschaften},
doi={10.22331/q-2019-03-05-128}
}

@article{gidney2024inplace,
  title={Inplace access to the surface code y basis},
  author={Gidney, Craig},
  journal={Quantum},
  volume={8},
  pages={1310},
  year={2024},
  publisher={Verein zur F{\"o}rderung des Open Access Publizierens in den Quantenwissenschaften},
doi={10.22331/q-2024-04-08-1310}
}

@misc{kornellsome,
      title={Some improvements to product formula circuits for Hamiltonian simulation}, 
      author={Andre Kornell and Peter Selinger},
      year={2025},
      eprint={2310.12256},
      archivePrefix={arXiv},
      primaryClass={quant-ph},
doi={10.48550/arXiv.2310.12256}
}

@misc{bartschi2024potential,
      title={Potential Applications of Quantum Computing at Los Alamos National Laboratory}, 
      author={Andreas Bärtschi and Francesco Caravelli and Carleton Coffrin and Jonhas Colina and Stephan Eidenbenz and Abhijith Jayakumar and Ammar A. Kirmani and Scott Lawrence and Minseong Lee and Andrey Y. Lokhov and Avanish Mishra and Sidhant Misra and Zachary Morrell and Zain Mughal and Duff Neill and Andrei Piryatinski and Allen Scheie and Marc Vuffray and Yu Zhang},
      year={2026},
      eprint={2406.06625},
      archivePrefix={arXiv},
      primaryClass={quant-ph},
doi={10.48550/arXiv.2406.06625}
}

@article{akahoshi2025compilation,
  title={Compilation of trotter-based time evolution for partially fault-tolerant quantum computing architecture},
  author={Akahoshi, Yutaro and Toshio, Riki and Fujisaki, Jun and Oshima, Hirotaka and Sato, Shintaro and Fujii, Keisuke},
  journal={PRX Quantum},
  volume={6},
  number={4},
  pages={040319},
  year={2025},
  publisher={APS},
doi={10.1103/93zr-1ykb}
}

@article{clark2021high,
  title={High-fidelity bell-state preparation with ca+ 40 optical qubits},
  author={Clark, Craig R and Tinkey, Holly N and Sawyer, Brian C and Meier, Adam M and Burkhardt, Karl A and Seck, Christopher M and Shappert, Christopher M and Guise, Nicholas D and Volin, Curtis E and Fallek, Spencer D and others},
  journal={Physical Review Letters},
  volume={127},
  number={13},
  pages={130505},
  year={2021},
  publisher={APS},
doi={10.1103/PhysRevLett.127.130505}
}

@misc{Obenland_pyLIQTR,
  author = {Obenland, Kevin and Elenewski, Justin and Morrell, Kaitlyn and Rempfer, Benjamin and Kuklinski, Parker and Neumann, Rylee Stuart and Kurlej, Arthur and Rood, Robert and Blue, John and Belarge, Joe},
  title = {pyLIQTR},
  howpublished = {GitHub repository},
  doi = {10.5281/zenodo.7221272},
  year = {2022},
  note = {MIT License}
}

@article{mcclean2020openfermion,
  title={OpenFermion: the electronic structure package for quantum computers},
  author={McClean, Jarrod R and Rubin, Nicholas C and Sung, Kevin J and Kivlichan, Ian D and Bonet-Monroig, Xavier and Cao, Yudong and Dai, Chengyu and Fried, E Schuyler and Gidney, Craig and Gimby, Brendan and others},
  journal={Quantum Science and Technology},
  volume={5},
  number={3},
  pages={034014},
  year={2020},
  publisher={IOP Publishing},
doi={10.1088/2058-9565/ab8ebc}
}

@misc{gidney2024magic,
      title={Magic state cultivation: growing T states as cheap as CNOT gates}, 
      author={Craig Gidney and Noah Shutty and Cody Jones},
      year={2024},
      eprint={2409.17595},
      archivePrefix={arXiv},
      primaryClass={quant-ph},
doi={10.48550/arXiv.2409.17595}
}

@article{litinski2019magic,
  title={Magic state distillation: Not as costly as you think},
  author={Litinski, Daniel},
  journal={Quantum},
  volume={3},
  pages={205},
  year={2019},
  publisher={Verein zur F{\"o}rderung des Open Access Publizierens in den Quantenwissenschaften},
doi={10.22331/q-2019-12-02-205}
}

@INPROCEEDINGS{hirano2025locality,
  author={Hirano, Yutaka and Fujii, Keisuke},
  booktitle={2025 IEEE International Conference on Quantum Computing and Engineering (QCE)}, 
  title={Locality-Aware Pauli-Based Computation for Local Magic State Preparation}, 
  year={2025},
  volume={01},
  number={},
  pages={670-680},
  doi={10.1109/QCE65121.2025.00078}}

@misc{haner2022space,
      title={Space-time optimized table lookup}, 
      author={Thomas Häner and Vadym Kliuchnikov and Martin Roetteler and Mathias Soeken},
      year={2022},
      eprint={2211.01133},
      archivePrefix={arXiv},
      primaryClass={quant-ph},
doi={10.48550/arXiv.2211.01133}
}

@article{molavi2025dependency,
  title={Dependency-aware compilation for surface code quantum architectures},
  author={Molavi, Abtin and Xu, Amanda and Tannu, Swamit and Albarghouthi, Aws},
  journal={Proceedings of the ACM on Programming Languages},
  volume={9},
  number={OOPSLA1},
  pages={57--84},
  year={2025},
  publisher={ACM New York, NY, USA},
doi={10.1145/3720416}
}

@inproceedings{silva2024multi,
  title={Multi-qubit Lattice Surgery Scheduling},
  author={Silva, Allyson and Zhang, Xiangyi and Webb, Zak and Kramer, Mia and Yang, Chan Woo and Liu, Xiao and Lemieux, Jessica and Chen, Ka-Wai and Scherer, Artur and Ronagh, Pooya},
  booktitle={19th Conference on the Theory of Quantum Computation, Communication and Cryptography (TQC 2024)},
  series={LIPIcs},
  volume={310},
  pages={1:1--1:22},
  year={2024},
  doi={10.4230/LIPIcs.TQC.2024.1},
}

@article{hamada2024efficient,
  title={Efficient and high-performance routing of lattice-surgery paths on three-dimensional lattice},
  author={Hamada, Kou and Suzuki, Yasunari and Tokunaga, Yuuki},
  journal={Quantum},
  volume={10},
  pages={2061},
  year={2026},
  publisher={Verein zur F{\"o}rderung des Open Access Publizierens in den Quantenwissenschaften},
doi={10.22331/q-2026-04-13-2061}
}

@misc{kan2025sparo,
      title={SPARO: Surface-code Pauli-based Architectural Resource Optimization for Fault-tolerant Quantum Computing}, 
      author={Shuwen Kan and Zefan Du and Chenxu Liu and Meng Wang and Yufei Ding and Ang Li and Ying Mao and Samuel Stein},
      year={2025},
      eprint={2504.21854},
      archivePrefix={arXiv},
      primaryClass={quant-ph},
doi={10.48550/arXiv.2504.21854}
}

@misc{litinski2022active,
      title={Active volume: An architecture for efficient fault-tolerant quantum computers with limited non-local connections}, 
      author={Daniel Litinski and Naomi Nickerson},
      year={2022},
      eprint={2211.15465},
      archivePrefix={arXiv},
      primaryClass={quant-ph},
doi={10.48550/arXiv.2211.15465}
}

@article{chamberland2022universal,
  title={Universal quantum computing with twist-free and temporally encoded lattice surgery},
  author={Chamberland, Christopher and Campbell, Earl T},
  journal={PRX Quantum},
  volume={3},
  number={1},
  pages={010331},
  year={2022},
  publisher={APS},
doi={10.1103/PRXQuantum.3.010331}
}

@misc{silva2024optimizing,
      title={Optimizing Multi-level Magic State Factories for Fault-Tolerant Quantum Architectures}, 
      author={Allyson Silva and Artur Scherer and Zak Webb and Abdullah Khalid and Bohdan Kulchytskyy and Mia Kramer and Kevin Nguyen and Xiangzhou Kong and Gebremedhin A. Dagnew and Yumeng Wang and Huy Anh Nguyen and Einar Gabbassov and Katiemarie Olfert and Pooya Ronagh},
      year={2025},
      eprint={2411.04270},
      archivePrefix={arXiv},
      primaryClass={quant-ph},
doi={10.48550/arXiv.2411.04270}
}

@techreport{coffrin2023quantum,
  title={Quantum Application Specifications and Benchmarks},
  author={Coffrin, Carleton and Morrell, Zachary},
  year={2023},
  institution={Los Alamos National Laboratory (LANL), Los Alamos, NM (United States)},
doi={10.11578/DC.20230922.2}
}

@misc{viszlai2023architecture,
      title={An Architecture for Improved Surface Code Connectivity in Neutral Atoms}, 
      author={Joshua Viszlai and Sophia Fuhui Lin and Siddharth Dangwal and Jonathan M. Baker and Frederic T. Chong},
      year={2023},
      eprint={2309.13507},
      archivePrefix={arXiv},
      primaryClass={quant-ph},
doi={10.48550/arXiv.2309.13507}
}

@article{skoric2023parallel,
  title={Parallel window decoding enables scalable fault tolerant quantum computation},
  author={Skoric, Luka and Browne, Dan E and Barnes, Kenton M and Gillespie, Neil I and Campbell, Earl T},
  journal={Nature Communications},
  volume={14},
  number={1},
  pages={7040},
  year={2023},
  publisher={Nature Publishing Group UK London},
doi={10.1038/s41467-023-42482-1}
}

@article{lin2025spatially,
  title={Spatially parallel decoding for multi-qubit lattice surgery},
  author={Lin, Sophia Fuhui and Peterson, Eric C and Sankar, Krishanu and Sivarajah, Prasahnt},
  journal={Quantum Science and Technology},
  volume={10},
  number={3},
  pages={035007},
  year={2025},
  publisher={IOP Publishing},
doi={10.1088/2058-9565/adc6b6}
}

@article{cain2024correlated,
  title={Correlated decoding of logical algorithms with transversal gates},
  author={Cain, Madelyn and Zhao, Chen and Zhou, Hengyun and Meister, Nadine and Ataides, J Pablo Bonilla and Jaffe, Arthur and Bluvstein, Dolev and Lukin, Mikhail D},
  journal={Physical Review Letters},
  volume={133},
  number={24},
  pages={240602},
  year={2024},
  publisher={APS},
doi={10.1103/PhysRevLett.133.240602}
}

@article{moflic2024constant,
  title={On the constant depth implementation of Pauli exponentials},
  author={Moflic, Ioana and Paler, Alexandru},
  journal={npj Quantum Information},
  year={2026},
  publisher={Nature Publishing Group UK London},
doi={10.1038/s41534-026-01226-x}
}

@article{geher2024error,
  title={Error-corrected Hadamard gate simulated at the circuit level},
  author={Geh{\'e}r, Gy{\"o}rgy P and McLauchlan, Campbell and Campbell, Earl T and Moylett, Alexandra E and Crawford, Ophelia},
  journal={Quantum},
  volume={8},
  pages={1394},
  year={2024},
  publisher={Verein zur F{\"o}rderung des Open Access Publizierens in den Quantenwissenschaften},
doi={10.22331/q-2024-07-02-1394}
}

@article{leblond2024realistic,
  title={Realistic cost to execute practical quantum circuits using direct clifford+ t lattice surgery compilation},
  author={LeBlond, Tyler and Dean, Christopher and Watkins, George and Bennink, Ryan},
  journal={ACM Transactions on Quantum Computing},
  volume={5},
  number={4},
  pages={1--28},
  year={2024},
  publisher={ACM New York, NY},
doi={10.1145/3689826}
}

@article{beverland2022surface,
  title={Surface code compilation via edge-disjoint paths},
  author={Beverland, Michael and Kliuchnikov, Vadym and Schoute, Eddie},
  journal={PRX Quantum},
  volume={3},
  number={2},
  pages={020342},
  year={2022},
  publisher={APS},
doi={10.1103/PRXQuantum.3.020342}
}

@article{watkins2023high,
  title={A High Performance Compiler for Very Large Scale Surface Code Computations},
  author={Watkins, George and Nguyen, Hoang Minh and Watkins, Keelan and Pearce, Steven and Lau, Hoi-Kwan and Paler, Alexandru},
  journal={Quantum},
  volume={8},
  pages={1354},
  year={2024},
  doi={10.22331/q-2024-05-22-1354}
}

@article{acharya2024quantum,
  title={Quantum error correction below the surface code threshold},
  author={Acharya, Rajeev and Abanin, Dmitry A and Aghababaie-Beni, Laleh and Aleiner, Igor and Andersen, Trond I and Ansmann, Markus and Arute, Frank and Arya, Kunal and Asfaw, Abraham and Astrakhantsev, Nikita and others},
  journal={Nature},
  year={2024},
doi={10.1038/s41586-024-08449-y}
}

@article{google2023suppressing,
  title={Suppressing quantum errors by scaling a surface code logical qubit},
  author={{Google Quantum AI}},
  journal={Nature},
  volume={614},
  number={7949},
  pages={676--681},
  year={2023},
  publisher={Nature Publishing Group UK London},
doi={10.1038/s41586-022-05434-1}
}

@article{zhao2022realization,
  title={Realization of an error-correcting surface code with superconducting qubits},
  author={Zhao, Youwei and Ye, Yangsen and Huang, He-Liang and Zhang, Yiming and Wu, Dachao and Guan, Huijie and Zhu, Qingling and Wei, Zuolin and He, Tan and Cao, Sirui and others},
  journal={Physical Review Letters},
  volume={129},
  number={3},
  pages={030501},
  year={2022},
  publisher={APS},
doi={10.1103/PhysRevLett.129.030501}
}

@article{bluvstein2024logical,
  title={Logical quantum processor based on reconfigurable atom arrays},
  author={Bluvstein, Dolev and Evered, Simon J and Geim, Alexandra A and Li, Sophie H and Zhou, Hengyun and Manovitz, Tom and Ebadi, Sepehr and Cain, Madelyn and Kalinowski, Marcin and Hangleiter, Dominik and others},
  journal={Nature},
  volume={626},
  number={7997},
  pages={58--65},
  year={2024},
  publisher={Nature Publishing Group UK London},
doi={10.1038/s41586-023-06927-3}
}

@misc{ruan2025trapsimd,
      title={TrapSIMD: SIMD-Aware Compiler Optimization for 2D Trapped-Ion Quantum Machines}, 
      author={Jixuan Ruan and Hezi Zhang and Xiang Fang and Ang Li and Wesley C. Campbell and Eric Hudson and David Hayes and Hartmut Haeffner and Travis Humble and Jens Palsberg and Yufei Ding},
      year={2025},
      eprint={2504.17886},
      archivePrefix={arXiv},
      primaryClass={quant-ph},
doi={10.48550/arXiv.2504.17886}
}

@article{grans2024improved,
  title={Improved pairwise measurement-based surface code},
  author={Grans-Samuelsson, Linnea and Mishmash, Ryan V and Aasen, David and Knapp, Christina and Bauer, Bela and Lackey, Brad and da Silva, Marcus P and Bonderson, Parsa},
  journal={Quantum},
  volume={8},
  pages={1429},
  year={2024},
  publisher={Verein zur F{\"o}rderung des Open Access Publizierens in den Quantenwissenschaften},
doi={10.22331/q-2024-08-02-1429}
}

@inproceedings{yin2025flexion,
  title={iSwitch: QEC on Demand via In-Situ Encoding of Bare Qubits for Ion Trap Architectures},
  author={Yin, Keyi and Fang, Xiang and Chen, Zhuo and Hayes, David and Kaur, Eneet and Nejabati, Reza and Haeffner, Hartmut and Campbell, Wes and Hudson, Eric and Palsberg, Jens and others},
  booktitle={Proceedings of the 31st ACM International Conference on Architectural Support for Programming Languages and Operating Systems, Volume 2},
  pages={1007--1021},
  year={2026},
doi={10.1145/3779212.3790177}
}

@inproceedings{leblond2023tiscc,
  title={Tiscc: A surface code compiler and resource estimator for trapped-ion processors},
  author={LeBlond, Tyler and Bennink, Ryan S and Lietz, Justin G and Seck, Christopher M},
  booktitle={Proceedings of the SC'23 Workshops of The International Conference on High Performance Computing, Network, Storage, and Analysis},
  pages={1426--1435},
  year={2023},
doi={10.1145/3624062.3624214}
}

@article{fowler2012surface,
  title={Surface codes: Towards practical large-scale quantum computation},
  author={Fowler, Austin G and Mariantoni, Matteo and Martinis, John M and Cleland, Andrew N},
  journal={Physical Review A},
  volume={86},
  number={3},
  pages={032324},
  year={2012},
  publisher={APS},
doi={10.1103/PhysRevA.86.032324}
}

@misc{beverland2022assessing,
      title={Assessing requirements to scale to practical quantum advantage}, 
      author={Michael E. Beverland and Prakash Murali and Matthias Troyer and Krysta M. Svore and Torsten Hoefler and Vadym Kliuchnikov and Guang Hao Low and Mathias Soeken and Aarthi Sundaram and Alexander Vaschillo},
      year={2022},
      eprint={2211.07629},
      archivePrefix={arXiv},
      primaryClass={quant-ph},
doi={10.48550/arXiv.2211.07629}
}

@article{terhal2015quantum,
  title={Quantum error correction for quantum memories},
  author={Terhal, Barbara M},
  journal={Reviews of Modern Physics},
  volume={87},
  number={2},
  pages={307--346},
  year={2015},
  publisher={APS},
doi={10.1103/RevModPhys.87.307}
}

@article{blunt2024compilation,
  title={Compilation of a simple chemistry application to quantum error correction primitives},
  author={Blunt, Nick S and Geh{\'e}r, Gy{\"o}rgy P and Moylett, Alexandra E},
  journal={Physical Review Research},
  volume={6},
  number={1},
  pages={013325},
  year={2024},
  publisher={APS},
doi={10.1103/PhysRevResearch.6.013325}
}

@misc{fowler2018low,
      title={Low overhead quantum computation using lattice surgery}, 
      author={Austin G. Fowler and Craig Gidney},
      year={2019},
      eprint={1808.06709},
      archivePrefix={arXiv},
      primaryClass={quant-ph},
doi={10.48550/arXiv.1808.06709}
}

\onecolumn
\appendix

\section{Resource Analysis for Clifford+T Circuits Directly Compiled to Lattice Surgery}
\label{sec:re_clifft}
In this Appendix, we provide a summary of the approach to direct Clifford+T lattice surgery compilation used in the present work as well as a summary of how the outputs of compilation are used to estimate the resource requirements of quantum circuits. As mentioned in Section~\ref{sec:direct} of the main text, our approach to direct Clifford+T lattice surgery compilation closely follows that of Ref.~\cite{leblond2024realistic} with minor modifications resulting from recent advances in the literature such as in-place Y state initialization~\cite{gidney2024inplace}. Also, as discussed in Section~\ref{sec:subcircuit}, the present work has generalized the approach to resource analysis from Ref.~\cite{leblond2024realistic} to handle circuits consisting of repetitions of subcircuits. As such, in this Appendix, we provide a primer for how resource analysis is performed for individual subcircuits in the present study, which follows the approach for whole circuits introduced in Ref.~\cite{leblond2024realistic}. 

\begin{figure*}[t]
    \centering
    \includegraphics[width=\textwidth]{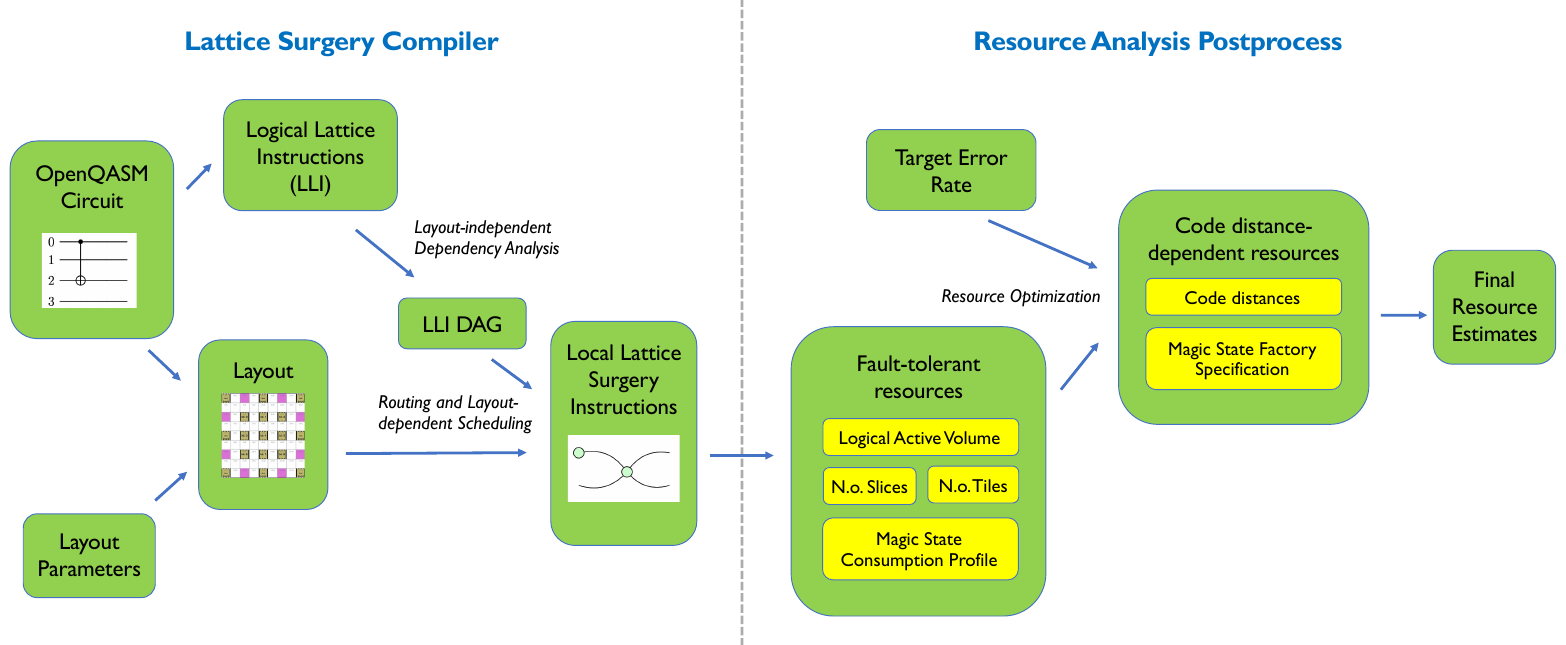}
    \captionsetup{skip=10pt}
    \caption{High-level overview of the Lattice Surgery Compiler (LSC) and its integration with our resource estimation methodology. Figure and caption re-used from Ref.~\cite{leblond2024realistic}.}
    \label{fig:pipeline}
\end{figure*}

For a schematic overview of the framework introduced in Ref.~\cite{leblond2024realistic}, see Fig.~\ref{fig:pipeline}, which depicts the division of the framework into two sub-processes. First, a lattice surgery compilation sub-process produces a circuit of primitive surface code instructions from an input logical circuit expressed in the Clifford+T gate set. This output consists of \textit{local} instructions (acting on at most one or two layout tiles), as well as magic state requests, organized into a sequence of time slices. Second, a resource analysis post-process reads statistics such as the total number of compiled slices, the total number of layout tiles, the logical active volume, and a magic state consumption profile (defined below) and determines the amount of space allocated to magic state factories and the magic state storage profile. As part of this process, the total computational space-time volume (for the whole circuit, not a subcircuit) is optimized by varying code distances under the constraint that the total logical error remains beneath a user-set threshold. When dealing with circuits composed of subcircuits, this optimization is performed with respect to the resource and logical error definitions from Section~\ref{sec:subcircuit}.

In what follows, we first highlight the motivation for a compiler-based resource estimation workflow and key insights gained from Ref.~\cite{leblond2024realistic}, and then we summarize important aspects of each of the sub-processes described above. For a more comprehensive discussion, see Ref.~\cite{leblond2024realistic}. 

\subsection{Motivation and Key Insights from Ref.~\cite{leblond2024realistic}}
The core motivation for including a lattice surgery compilation sub-process within a resource estimation framework for surface code quantum computers is to more realistically estimate the resource requirements for executing circuits by (i) explicitly translating parallelism from the input circuit into parallelism of fault-tolerance-layer operations, (ii) obtaining more accurate logical error rates by decomposing long-range gates into pair-wise local lattice surgery primitives and counting the \textit{active} computational volumes of each, and (iii) using the ``real-time'' magic state consumption profile of the circuit to optimize resources dedicated to magic state production and storage.

More specifically, computation of the (layout-dependent) number of compiled time slices of parallelized fault-tolerance-layer operations from input quantum circuits enables the sharp computation of their execution times without forcing the use of a compilation scheme that makes time-to-solution estimates easier to calculate by, e.g., serializing the computation (as is the done in SPBC). Also, though not emphasized in the present work, the computation of a realistic logical ``active volume'' enables a more realistic logical error calculation than is often done by, e.g., multiplying a uniform logical error rate by the total space-time volume~\cite{beverland2022assessing}. Notably, the framework in Ref.~\cite{leblond2024realistic} is end-to-end in the sense that the output instruction set is a set of local, layout-independent fault-tolerant surface code instructions that may each have their own resource requirements and logical error rates; for an example of a compiler that we designed to translate each member of our local instruction set into logically-verified native hardware circuits and count resource costs for them, see Ref.~\cite{leblond2023tiscc}\footnote{Although Ref.~\cite{leblond2024realistic} and the present work do not make use of the flexibility to define separate logical error rates for each primitive instruction, it is significant that the framework allows for it, as incorporating those details could substantially improve the realism of resource estimates.}. 

Significantly, our analysis of the magic state production-storage trade space in Ref.~\cite{leblond2024realistic} (using the magic state consumption profile output by compilation) led us to the conclusion that allocating a minimal, constant number of continuously operating magic state factories was highly detrimental to resource overheads due to the significant cost of storing magic states, which results from the requirement that magic state demand is satisfied at all points of the computation and the practical circumstance that magic state consumption is variable throughout circuit execution. We discovered that this issue could be resolved by allocating space for the number of factories required by the highest-consumption portion of the circuit and dynamically toggling factories off and on, according to real-time demand, to minimize storage requirements. This strategy, called \textit{min-storage}, was found to be optimal from a space-time volume standpoint (even though it notably over-provisions magic state factories) because the strategy that minimizes the number of allocated factories inevitably causes huge build-ups of stored magic states during lower-consumption portions of the circuit.

Although the framework in Ref.~\cite{leblond2024realistic} was designed to advance the realism of the state-of-the-art from works such as Ref.~\cite{beverland2022assessing}, it still contained certain limitations. For example, it avoided dealing with dynamic behavior by guaranteeing a high magic state factory success rate ($\simeq 99\%$) and not handling any potential failures. Also, space-time volume was allocated for every conditional $S$ gate following $T$ injection in order to make compilation static. This latter compromise was made, not because of any limitation of the LSC (the original version of the compiler implemented a gate streaming pipeline), but in order to enable greater optimization of parallelism and resource state management in the fault-tolerance layer. Neither of these limitations were improved upon in the present study.

\subsection{Lattice Surgery Compilation}
Ref.~\cite{leblond2024realistic} presented a number of improvements to and benchmarking results for the open-source Lattice Surgery Compiler (LSC) (originally introduced in Ref.~\cite{watkins2023high}). The upgraded Lattice Surgery Compiler (LSC) operates in two distinct stages of translation:
\begin{enumerate}
  \item \textbf{Logical Clifford+T Circuit to Logical Lattice Instructions (LLI).} Each logical gate (Clifford+T gate set) is converted into a sequence of layout-independent intermediate instructions, called \emph{logical lattice instructions} (LLI). This step occurs prior to routing and ordering of operations into time slices. See Table~\ref{tab:lli} for a summary of the input logical instruction set and the LLI members that each is translated into, as well as the nominal footprint of each gate. The result of this stage of compilation is a directed acyclic graph (DAG) of LLI.
  \item \textbf{Scheduling, Routing, and Decomposition into Local Instructions.} LLI are scheduled according to the wave pipeline scheduling algorithm presented in Ref.~\cite{leblond2024realistic}, which is an As-Soon-As-Possible scheduler that accounts for resource conflicts (routing tile and magic state availability) while respecting constraints set by a user-specified surface code layout. Upon scheduling, applicable LLI are decomposed into a \textit{local} instruction set acting on pairs of layout tiles for a specified number of slices (see Table~\ref{tab:local_instructions} for a list of primitives used in the decomposition of the BellBasedCNOT LLI). Our decomposition strategy for CNOT gates follows the one based on long-range Bell states from Ref.~\cite{beverland2022surface}. We note that, though we have implemented the edge-disjoint path compilation strategy from Ref.~\cite{beverland2022surface} within the LSC in its own pipeline (separate from the wave pipeline), its benefits on practical circuits have not been carefully evaluated. The wave pipeline, used in the present work, greedily schedules BellBasedCNOT LLI by allocating vertex-disjoint paths based on an A* heuristic. Random circuit benchmarking results in Ref.~\cite{leblond2024realistic} confirm that slice counts scale consistently with worst-case EDPC expectations (e.g., $\Omega(\sqrt{m})$ depth for $m$ parallel CNOTs).
\end{enumerate}

The output of the LSC is a slice-by-slice (line-by-line) list of comma-separated LLI (including resource state requests) with their decomposition into local instructions given in brackets. Statistics such as slice count and active volume are also output by the LSC. From the time-sliced output file, a magic state consumption profile can easily be extracted for use in the resource optimization post-process.

\begin{table*}[t]
  \centering
  \caption{Logical instruction set used in this work together with the logical lattice instructions (LLI) involved in the compilation of each instruction. Where possible, we show the numbers of tiles and time slices that each gate is ultimately compiled to. Wherever BellBasedCNOTs are involved, the number of tiles is route-dependent. The details given for  T/$\mathrm{T}^{\dagger}$ gates do not include those of the required corrective S gate. Table and caption adapted from Ref.~\cite{leblond2024realistic}.}
  \begin{tabular}{|l|p{8cm}|c|}
    \hline
    \textbf{Logical Gate} & \textbf{Involved Logical Lattice Instructions (LLI)} & \textbf{N.o. Tiles, N.o. Slices} \\
    \hline
Reset & Reset &  0, 0 \\
X, Y, Z & XGate, YGate, ZGate & 0, 0 \\
H & HGate, RotateSingleCellPatch & 2, 2 \\
CNOT & BellBasedCNOT & Route-dependent, 2 \\
T/$\mathrm{T}^{\dagger}$ & MagicStateRequest, BellBasedCNOT, SinglePatchMeasurement & Route-dependent, 2 \\
S/$\mathrm{S}^{\dagger}$ & YStateRequest, MultiPatchMeasurement, SinglePatchMeasurement & Route-dependent, 2 \\
    \hline
  \end{tabular}
  \label{tab:lli}
\end{table*}

\begin{table*}[t]
  \centering
  \caption{Local instruction set implemented within the second stage of lattice surgery compilation in the LSC to decompose the BellBasedCNOT LLI. Table and caption adapted from Ref.~\cite{leblond2024realistic}.}
  \scalebox{0.9}{
  \begin{tabular}{|l|p{7cm}|c|c|}
    \hline
    \textbf{Instruction} & \textbf{Description} & \textbf{Tiles In/Out} & \textbf{Logical Time Slices} \\
    \hline
    TwoPatchMeasure& Measures the joint XX/ZZ operators of two vertically/horizontally-adjacent initialized tiles & 2 & 1 \\
    BellPrepare & Initializes a Bell state on two adjacent uninitialized tiles & 2 & 1 \\
    BellMeasure & Performs a destructive Bell basis measurement on two adjacent (initialized) tiles and makes uninitialized & 2 & 1 \\
    ExtendSplit & A patch extension followed by a split operation & 2 & 1 \\
    MergeContract& A merge operation followed by a patch contraction & 2 & 1 \\
    Move & A patch extension followed by a patch contraction & 2 & 1 \\
    \hline
  \end{tabular}
  }
  \label{tab:local_instructions}
\end{table*}

\subsection{Resource Analysis Post-Process}

With the aforementioned outputs of compilation, we employ a resource analysis post-process to optimize total space-time volume subject to a target logical error. At this stage, abstract FTRE such as the number of logical time slices $\tau_{\mathrm{logical}}$, the number of tiles in the logical block $n_{\mathrm{logical}}$, and the active volume of logical computation $V_{\mathrm{logical}}$ have been completely determined by compilation. What remains is to determine the same contributions from magic state distillation and magic state storage, as well as the minimum allowed code distance(s) to respect a target total logical error. We consider a simple additive resource model: the total number of time slices is $\tau_{\mathrm{total}} = \tau_{\mathrm{logical}} + w \tau_D$, where $w$ is a number of warm-up distillation cycles and $\tau_D$ is the number of time slices required by a single distillation cycle. Similarly, the total spatial footprint is $n_{\mathrm{total}} = n_{\mathrm{logical}} + n_{\mathrm{storage}} + N n_D$, where $N$ is the number of allocated magic state factories and $n_D$ is the number of logical tiles required by a magic state factory. Finally, the total logical error is $\epsilon_{\mathrm{total}} = \epsilon_{\mathrm{logical}} + \epsilon_{\mathrm{dist}} + \epsilon_{\mathrm{storage}}$. We will return to the calculation of these error contributions from active volumes after a digression on magic state factory specification and \textit{min-storage}.

The magic state distillation factory selection framework assumed by the present study relies heavily on the factory selection framework described in Ref.~\cite{beverland2022assessing} but makes several additional assumptions in accordance with the description in Ref.~\cite{leblond2024realistic}. Briefly, two rounds of 15:1 distillation are used, with generically two different code distances $d_1$ and $d_2$. The number of level 1 distillation units is chosen to ensure that the total acceptance rate is greater than a pre-determined threshold ($\simeq 99 \%$). The resource overheads of each distillation factory assume a particular compilation of the 15:1 distillation circuit into the same set of local lattice surgery primitives targeted by the LSC (e.g., those found in Table~\ref{tab:local_instructions}). See Fig. 9 and Table VI of Ref.~\cite{beverland2022assessing} for further details of this compilation as well as its corresponding resource overheads. As mentioned in the main text, the logical error rates for distilled states required by the circuits under consideration in this study are too low for modern protocols like cultivation to be considered practical~\cite{gidney2024magic}. We treat factory selection as a function that takes in $d_1$ and $d_2$ and outputs $n_D$, $\tau_D$, and $P_T$, the logical error rate of a distilled magic state.

As alluded to, the resource analysis post-process requires that magic states are produced and stored such that a sufficient quantity of magic states is available to supply every consumption cycle. In essence, the LSC compiles the circuit statically assuming magic state availability at certain tiles given a user-set rate of replenishment (here taken to be a single logical time slice), while the resource analysis post-process specifies the factory and determines how many are needed to meet the demand. Knowing $\tau_D$, the magic state consumption profile is calculated according to a coarse-graining by distillation cycle. The number of magic states consumed through distillation cycle $k$ is $m(k) \equiv \sum_{i=1}^{k\tau_D} m_i$, where $k \tau_D$ is the number of time slices required to complete $k$ distillation cycles and $m_i$ is the number of magic state requests found in slice $i$ of the compiled output (i.e. the number of \verb+MagicStateRequest+ LLI on line $i$ of the sliced output file of the LSC). While Ref.~\cite{leblond2024realistic} explored several approaches to meeting magic state demand, \textit{min-storage} yielded the most optimal space-time volume, and hence, \textit{min-storage} is the only approach we will focus on here. In $\textit{min-storage}$, the number of magic state factories active during distillation cycle $k$ is $N(k) = m(k+1) - m(k)$, where $N(0) = m(1)$ defines the number of active factories during a single warm-up cycle (i.e. we choose $w = 1$ in the above expression for $\tau_{\mathrm{total}}$). The number of magic state factories allocated during program execution is $N \equiv N_{\mathrm{max}} = \max_k N(k)$. Significantly, there are unused distillation tiles whenever $N(k) \neq N_{\mathrm{max}}$. Similarly, we approximate the number of magic states in storage from distillation cycle $k$ as $R(k) = N(k)$, reserve $n_{\mathrm{storage}} = \max_k R(k)$ tiles for stored magic states, and approximate the storage active volume as $V_{\mathrm{storage}} \simeq \tau_D\sum_{k=0}^{k_{\mathrm{max}} - 1}R(k)$.

 Altogether, we calculate $\epsilon_{\mathrm{logical}} = V_{\mathrm{logical}} P(d)$ and $\epsilon_{\mathrm{storage}} = V_{\mathrm{storage}} P(d)$, where $P(d) = 0.1 d \left(\frac{p}{0.01}\right)^{\frac{d+1}{2}}$~\cite{fowler2018low} and $p = 6\times 10^{-4}$\cite{clark2021high} are reasonable yet optimistic assumptions on logical and physical error rates. The distillation contribution to error is $\epsilon_{\mathrm{dist}} = m(k_{\mathrm{max}})P_T$, where $m(k_{\mathrm{max}})$ is the total number of consumed magic states. We note that, although separate logical error rates for each instruction can in principle be utilized to obtain a more realistic total error (an active volume for each instruction can easily be derived from the compiler output), we have opted to use the uniform logical error rate above for the whole active volume.

\section{Detailed Fault-Tolerant Resource Estimates}
\label{sec:tables}

\begin{table*}
\caption{Fault-tolerance-layer resource estimates (FTRE) in the SPBC scheme. See Sec.~\ref{sec:spbc} for specification of scheme and Sec.~\ref{sec:ftre} for definitions of the FTRE we consider.}
\centering
\scalebox{0.65}{
\begin{tabular}{|l|l|rrrr|rrrr|rrrr|}
\toprule
Application & Algorithm & $d_1$ & $d_2$ & $N$ & $P_T$ & $n_{\mathrm{total}}$ & $\tau_{\mathrm{total}}$ & $\tau_{\mathrm{total}} \times d_2$ & $n_{\mathrm{total}} \times \tau_{\mathrm{total}} \times d_2^3$ & $\epsilon_{\mathrm{logical}}$ & $\epsilon_{\mathrm{dist}}$ & $\epsilon_{\mathrm{storage}}$ & $\epsilon_{\mathrm{total}}$ \\
\midrule
TFIM (square) & Trotter & 11 & 31 & 3 & 2.37e-18 & 3.96e+02 & 3.70e+13 & 1.15e+15 & 4.36e+20 & 6.59e-04 & 7.30e-06 & 9.70e-06 & 6.76e-04 \\
TFIM (triangle) & Trotter & 11 & 35 & 3 & 1.75e-18 & 2.20e+03 & 2.11e+15 & 7.39e+16 & 1.99e+23 & 1.54e-03 & 3.08e-04 & 2.25e-06 & 1.85e-03 \\
Kitaev (triangle) & Trotter & 11 & 37 & 3 & 1.75e-18 & 2.19e+03 & 3.31e+16 & 1.22e+18 & 3.66e+24 & 1.53e-03 & 4.82e-03 & 2.24e-06 & 6.36e-03 \\
TFIM (cube) & Trotter & 11 & 35 & 3 & 1.75e-18 & 3.61e+03 & 4.53e+15 & 1.58e+17 & 7.01e+23 & 5.57e-03 & 6.61e-04 & 4.83e-06 & 6.24e-03 \\
Kitaev (honeycomb) & Trotter & 11 & 37 & 3 & 1.75e-18 & 4.49e+03 & 1.81e+16 & 6.71e+17 & 4.12e+24 & 1.78e-03 & 2.64e-03 & 1.23e-06 & 4.43e-03 \\
$\mathrm{\alpha-RuCl_3}$ & Trotter & 11 & 37 & 3 & 1.75e-18 & 4.49e+03 & 8.41e+15 & 3.11e+17 & 1.91e+24 & 8.26e-04 & 1.23e-03 & 5.69e-07 & 2.05e-03 \\
\hline
TFIM (square) & QSP & 9 & 27 & 3 & 4.45e-15 & 4.10e+02 & 1.88e+11 & 5.08e+12 & 1.52e+18 & 9.63e-04 & 6.98e-05 & 1.19e-05 & 1.04e-03 \\
TFIM (triangle) & QSP & 11 & 31 & 3 & 2.37e-18 & 2.29e+03 & 4.66e+13 & 1.44e+15 & 3.18e+21 & 8.56e-03 & 9.20e-06 & 1.22e-05 & 8.58e-03 \\
Kitaev (triangle) & QSP & 11 & 31 & 3 & 2.37e-18 & 2.29e+03 & 4.32e+13 & 1.34e+15 & 2.96e+21 & 7.95e-03 & 8.54e-06 & 1.13e-05 & 7.97e-03 \\
TFIM (cube) & QSP & 11 & 33 & 3 & 1.79e-18 & 3.68e+03 & 5.61e+13 & 1.85e+15 & 7.42e+21 & 1.10e-03 & 8.36e-06 & 9.39e-07 & 1.11e-03 \\
Kitaev (honeycomb) & QSP & 11 & 33 & 3 & 1.79e-18 & 4.57e+03 & 4.75e+13 & 1.57e+15 & 7.81e+21 & 1.17e-03 & 7.09e-06 & 7.97e-07 & 1.18e-03 \\
$\mathrm{\alpha-RuCl_3}$ & QSP & 11 & 35 & 3 & 1.75e-18 & 4.56e+03 & 2.49e+15 & 8.71e+16 & 4.87e+23 & 3.91e-03 & 3.64e-04 & 2.66e-06 & 4.27e-03 \\
\bottomrule
\end{tabular}
}
\label{tab:FTRE1}
\end{table*}

\begin{table*}
\caption{Fault-tolerance-layer resource estimates (FTRE) in the direct Clifford+T compilation scheme using the minimal-storage approach to resource optimization and the 1-lane-condensed layout from Ref.~\cite{leblond2024realistic}. See Sec.~\ref{sec:direct} and~\ref{sec:subcircuit} for further details of scheme and Sec.~\ref{sec:ftre} for definitions of the FTRE we consider.}
\centering
\scalebox{0.65}{
\begin{tabular}{|l|l|rrrr|rrrr|rrrr|}
\toprule
Application & Algorithm & $d_1$ & $d_2$ & $N$ & $P_T$ & $n_{\mathrm{total}}$ & $\tau_{\mathrm{total}}$ & $\tau_{\mathrm{total}} \times d_2$ & $n_{\mathrm{total}} \times \tau_{\mathrm{total}} \times d_2^3$ & $\epsilon_{\mathrm{logical}}$ & $\epsilon_{\mathrm{dist}}$ & $\epsilon_{\mathrm{storage}}$ & $\epsilon_{\mathrm{total}}$ \\
\midrule
TFIM (square) & Trotter & 11 & 29 & 92 & 1.14e-17 & 7.04e+03 & 1.08e+12 & 3.12e+13 & 1.85e+20 & 2.68e-04 & 3.52e-05 & 7.56e-05 & 3.79e-04 \\
TFIM (triangle) & Trotter & 11 & 31 & 293 & 2.37e-18 & 2.14e+04 & 2.65e+13 & 8.23e+14 & 1.69e+22 & 3.55e-03 & 4.17e-04 & 2.77e-04 & 4.25e-03 \\
Kitaev (triangle) & Trotter & 11 & 33 & 292 & 1.79e-18 & 1.90e+04 & 3.90e+14 & 1.29e+16 & 2.66e+23 & 3.33e-03 & 4.93e-03 & 2.77e-04 & 8.54e-03 \\
TFIM (cube) & Trotter & 11 & 33 & 393 & 1.79e-18 & 2.64e+04 & 4.97e+13 & 1.64e+15 & 4.71e+22 & 6.89e-04 & 6.75e-04 & 3.79e-05 & 1.40e-03 \\
Kitaev (honeycomb) & Trotter & 11 & 33 & 435 & 1.79e-18 & 2.98e+04 & 1.84e+14 & 6.06e+15 & 1.97e+23 & 3.12e-03 & 2.70e-03 & 1.52e-04 & 5.98e-03 \\
$\mathrm{\alpha-RuCl_3}$ & Trotter & 11 & 33 & 449 & 1.79e-18 & 3.06e+04 & 8.70e+13 & 2.87e+15 & 9.57e+22 & 1.48e-03 & 1.25e-03 & 7.05e-05 & 2.81e-03 \\
\hline
TFIM (square) & QSP & 9 & 25 & 11 & 6.62e-15 & 1.12e+03 & 1.92e+11 & 4.81e+12 & 3.35e+18 & 7.58e-03 & 1.04e-04 & 9.21e-05 & 7.81e-03 \\
TFIM (triangle) & QSP & 11 & 31 & 13 & 2.37e-18 & 3.64e+03 & 4.58e+13 & 1.42e+15 & 4.97e+21 & 4.22e-03 & 9.20e-06 & 6.11e-06 & 4.23e-03 \\
Kitaev (triangle) & QSP & 11 & 31 & 13 & 2.37e-18 & 3.64e+03 & 4.05e+13 & 1.26e+15 & 4.39e+21 & 3.73e-03 & 8.54e-06 & 5.67e-06 & 3.75e-03 \\
TFIM (cube) & QSP & 11 & 33 & 15 & 1.79e-18 & 5.20e+03 & 7.05e+13 & 2.33e+15 & 1.32e+22 & 6.93e-04 & 8.36e-06 & 4.70e-07 & 7.02e-04 \\
Kitaev (honeycomb) & QSP & 11 & 31 & 14 & 2.37e-18 & 6.37e+03 & 4.50e+13 & 1.39e+15 & 8.54e+21 & 8.67e-03 & 9.39e-06 & 6.24e-06 & 8.69e-03 \\
$\mathrm{\alpha-RuCl_3}$ & QSP & 11 & 35 & 16 & 1.75e-18 & 6.28e+03 & 2.45e+15 & 8.57e+16 & 6.59e+23 & 1.92e-03 & 3.64e-04 & 1.33e-06 & 2.29e-03 \\
\bottomrule
\end{tabular}
}
\label{tab:FTRE2}
\end{table*}

\begin{table*}
\caption{Same as Table~\ref{tab:FTRE2} except using the 1-lane layout from Ref.~\cite{leblond2024realistic} instead of 1-lane-condensed.}
\centering
\scalebox{0.65}{
\begin{tabular}{|l|l|rrrr|rrrr|rrrr|}
\toprule
Application & Algorithm & $d_1$ & $d_2$ & $N$ & $P_T$ & $n_{\mathrm{total}}$ & $\tau_{\mathrm{total}}$ & $\tau_{\mathrm{total}} \times d_2$ & $n_{\mathrm{total}} \times \tau_{\mathrm{total}} \times d_2^3$ & $\epsilon_{\mathrm{logical}}$ & $\epsilon_{\mathrm{dist}}$ & $\epsilon_{\mathrm{storage}}$ & $\epsilon_{\mathrm{total}}$ \\
\midrule
TFIM (square) & Trotter & 11 & 27 & 143 & 1.52e-16 & 1.25e+04 & 6.63e+11 & 1.79e+13 & 1.64e+20 & 3.00e-03 & 4.68e-04 & 1.24e-03 & 4.70e-03 \\
TFIM (triangle) & Trotter & 11 & 31 & 426 & 2.37e-18 & 3.18e+04 & 1.61e+13 & 4.98e+14 & 1.52e+22 & 2.67e-03 & 4.17e-04 & 2.77e-04 & 3.36e-03 \\
Kitaev (triangle) & Trotter & 11 & 33 & 415 & 1.79e-18 & 2.77e+04 & 2.77e+14 & 9.13e+15 & 2.76e+23 & 2.84e-03 & 4.93e-03 & 2.77e-04 & 8.05e-03 \\
TFIM (cube) & Trotter & 11 & 31 & 544 & 2.37e-18 & 4.24e+04 & 3.24e+13 & 1.00e+15 & 4.09e+22 & 8.48e-03 & 8.95e-04 & 5.94e-04 & 9.97e-03 \\
Kitaev (honeycomb) & Trotter & 11 & 33 & 571 & 1.79e-18 & 4.14e+04 & 1.27e+14 & 4.20e+15 & 1.89e+23 & 2.57e-03 & 2.70e-03 & 1.52e-04 & 5.43e-03 \\
$\mathrm{\alpha-RuCl_3}$ & Trotter & 11 & 33 & 569 & 1.79e-18 & 4.13e+04 & 5.92e+13 & 1.96e+15 & 8.79e+22 & 1.21e-03 & 1.25e-03 & 7.05e-05 & 2.53e-03 \\
\hline
TFIM (square) & QSP & 9 & 25 & 11 & 6.62e-15 & 1.34e+03 & 1.78e+11 & 4.45e+12 & 3.73e+18 & 7.16e-03 & 1.04e-04 & 9.21e-05 & 7.38e-03 \\
TFIM (triangle) & QSP & 11 & 31 & 14 & 2.37e-18 & 5.66e+03 & 4.20e+13 & 1.30e+15 & 7.07e+21 & 4.01e-03 & 9.20e-06 & 6.11e-06 & 4.03e-03 \\
Kitaev (triangle) & QSP & 11 & 31 & 14 & 2.37e-18 & 5.66e+03 & 3.70e+13 & 1.15e+15 & 6.23e+21 & 3.54e-03 & 8.54e-06 & 5.67e-06 & 3.55e-03 \\
TFIM (cube) & QSP & 11 & 33 & 15 & 1.79e-18 & 8.41e+03 & 6.60e+13 & 2.18e+15 & 1.99e+22 & 6.52e-04 & 8.36e-06 & 4.70e-07 & 6.61e-04 \\
Kitaev (honeycomb) & QSP & 11 & 31 & 14 & 2.37e-18 & 1.03e+04 & 4.11e+13 & 1.27e+15 & 1.26e+22 & 7.95e-03 & 9.39e-06 & 6.24e-06 & 7.97e-03 \\
$\mathrm{\alpha-RuCl_3}$ & QSP & 11 & 35 & 16 & 1.75e-18 & 1.02e+04 & 2.24e+15 & 7.85e+16 & 9.82e+23 & 1.77e-03 & 3.64e-04 & 1.33e-06 & 2.13e-03 \\
\bottomrule
\end{tabular}
}
\label{tab:FTRE3}
\end{table*}

Tables \ref{tab:FTRE1} and \ref{tab:FTRE2} show detailed fault-tolerant resource estimates (FTRE) for the SPBC and direct Clifford+T lattice surgery compilation scheme, respectively, for the applications under consideration in this study. It is from these two tables that the results presented in Sec.~\ref{sec:results} were derived. Alternative FTRE for the direct Clifford+T lattice surgery compilation scheme using the 1-lane layout from Ref.~\cite{leblond2024realistic} instead of the 1-lane-condensed layout are presented in Table~\ref{tab:FTRE3}. It can be seen that the results are fairly comparable between Tables~\ref{tab:FTRE2} and~\ref{tab:FTRE3}, with the total footprint for most Trotter circuits slightly benefiting from the 1-lane layout and that of most QSP circuits slightly benefiting from the 1-lane-condensed layout.

\end{document}